\documentclass[%
prx,twocolumn,
notitlepage,
 amsmath,amssymb,
 aps,
prx,
floatfix,
longbibliography
]{revtex4-1}

\usepackage{graphicx}
\usepackage{dcolumn}
\usepackage{bm}
\usepackage{xcolor}
\usepackage{float}
\usepackage{bbold}
\usepackage{color}
\usepackage[normalem]{ulem}
\usepackage{esint}
\definecolor{bleuf}{rgb}{0,0.44,0.72}
\usepackage[unicode=true,pdfusetitle,bookmarks=true,
bookmarksnumbered=false,bookmarksopen=false,breaklinks=false,
pdfborder={0 0 0},backref=false,colorlinks=true,citecolor=bleuf,
linkcolor=bleuf,urlcolor=bleuf]{hyperref}

\newcommand\tr{\mathrm{tr}}

\allowdisplaybreaks[3]

\begin{document}

\title{Topology-driven ordering of flocking matter}
\author{Am\'elie Chardac$^1$}
\author{Ludwig A. Hoffmann$^2$}
\author{Yoann Poupart$^1$}
\author{Luca Giomi$^2$}
\author{Denis Bartolo$^1$}
\affiliation{$^1$Univ.~Lyon, ENS de Lyon, Univ.~Claude Bernard, CNRS, Laboratoire de Physique, F-69342, Lyon, France}
\affiliation{$^2$ Instituut-Lorentz, Universiteit Leiden, P.O. Box 9506, 2300 RA Leiden, The Netherlands}

%
\begin{abstract}
When interacting motile units  self-organize into flocks, they realize one of the most robust ordered state found in nature. However, after twenty five years of intense research, the very mechanism controlling the ordering dynamics of both living and artificial flocks has remained unsettled.
Here, combining active-colloid experiments, numerical simulations and analytical work, we explain how flocking liquids heal their spontaneous flows initially plagued by collections of topological defects to achieve long-ranged polar order even in two dimensions. 
We demonstrate that the self-similar ordering of flocking matter is ruled by a living network of domain walls linking all $\pm1$ vortices, and guiding their annihilation dynamics. 
Crucially, this singular orientational structure echoes the formation of extended density patterns in the shape of interconnected bow ties. We establish that this double structure emerges from the interplay between self-advection and density gradients dressing each $-1$ topological charges with four orientation walls.  We then explain how active Magnus forces link all topological charges with extended domain walls, while elastic interactions drive their attraction along the resulting filamentous network of polarization singularities. Taken together our experimental, numerical and analytical results illuminate the suppression of all flow singularities, and the emergence of pristine unidirectional order in flocking matter.
\end{abstract}

\maketitle

Dazzling nonequilibrium steady states are consistently observed in soft condensed matter assembled from motile units~\cite{Marchetti_review,GranickNew,Zottl2016,Doostmohammadi2018}.  
But  their lively dynamics comes at a high price. Unlike in equilibrium, the  interplay between the inner structure and flows of  active matter prohibits the emergence of macroscopic order~\cite{Doostmohammadi2018,Shankar_review,Saintillan2012}. At large scales, the  sustained proliferation of topological defects traps virtually all synthetic active materials in isotropic and chaotic dynamical states. This  picture of topological charges endlessly rampaging through active crystals and liquid crystals finds a remarkable exception in flocks~\cite{Toner_review,Cavagna_Review,Chate_review}. Flocks generically refers to collections of interacting polar units collectively moving along the same average direction~\cite{Toner_review} as observed over more than six orders of magnitude in scale, from kilometer-long insect swarms 
to  colloidal and molecular flocks cruising through microfluidic devices~\cite{Bausch2010,Dauchot2010,Sood2014,Bricard2013,Yan2016}. 
As a result of an intimate interplay between  orientational and density excitations captured by Toner--Tu hydrodynamics~\cite{Marchetti_review,Chate_review,Toner_review}, both living and synthetic flocks realize one of the most stable broken symmetry phase observed in nature, in vitro and in silico: flocks can support genuine long-ranged polar order both in three and two dimensions, even when challenged by thermal fluctuations or quenched isotropic disorder~\cite{Toner95,Peruani2013,Toner2018,Chardac2020,Chate_review}. 
However, after twenty five years of intense research, the very mechanism controlling the ordering dynamics of flocking matter remains unsettled. The question is deceptively simple: starting from a homogeneous ensemble of motile particles undergoing uncoordinated motion, how does their velocity field initially marred by a number of topological defects heal to reach pristine orientational order, as illustrated in Supplementary Movie 1 and Fig.~\ref{Fig1}? 

The response to this fundamental question remains elusive or restricted to idealized incompressible systems~\cite{Baskaran2010,Rana2020}, and the essential obstacle to elucidate the phase ordering of polar active matter lies in our poor understanding of their topological defects. Dues to the inherent coupling between polar order and density via self-propulsion, the topological defects of flocking matter  are highly non linear objects yielding non-local flow distortions and extended density perturbations~\cite{Kruse2004,Bricard2015,Shankar2017,Chardac2020,Chate_review}. 
  As a results, unlike in active nematics, or passive systems such as ferromagnets, superfluids, liquid crystals or even model universe~\cite{Bray2002,Chuang1991,Pargellis94,Bowick1994,Ruutu1996,Shankar2019,Giomi2013}, the very principles ruling the interactions and annihilation of  topological charges in flocks remain out of reach of our current understanding of active condensed matter.
  
In this article, we describe the elementary topological excitations of two dimensional flocking liquids, and elucidate their phase ordering dynamics. 
To do so, we first characterize the coarsening of colloidal-roller liquids after a rapid quench in the flock phase. 
We show that a self-similar dynamics emerges from the annihilation of $\pm 1$ vortices along a filamentous network of domain walls with no counterparts in passive materials. 
This lively orientational structure is mirrored by very characteristic density patterns having the shape of interconnected bow ties generic to all realizations  of Toner-Tu fluids. 
Combining experiments, numerics and theory, we establish that this double structure is determined by extended singularity lines growing from  $-1$ vortices, and  shaped by the competition between self-advection and pressure gradients. 
Finally, we model the annihilation dynamics of vortex pairs and show how the active analogue of  Magnus forces links defects of opposite topological charges with domain walls localizing all  shear deformations of the spontaneously flowing liquids. In turn, orientational elasticity attract all topological defects along this emergent filamentous structure which eventually vanishes to form a  material assembled from self-propelled units all flocking along the same direction. 
\begin{figure*}
	\begin{center}
	\includegraphics[width=1\textwidth]{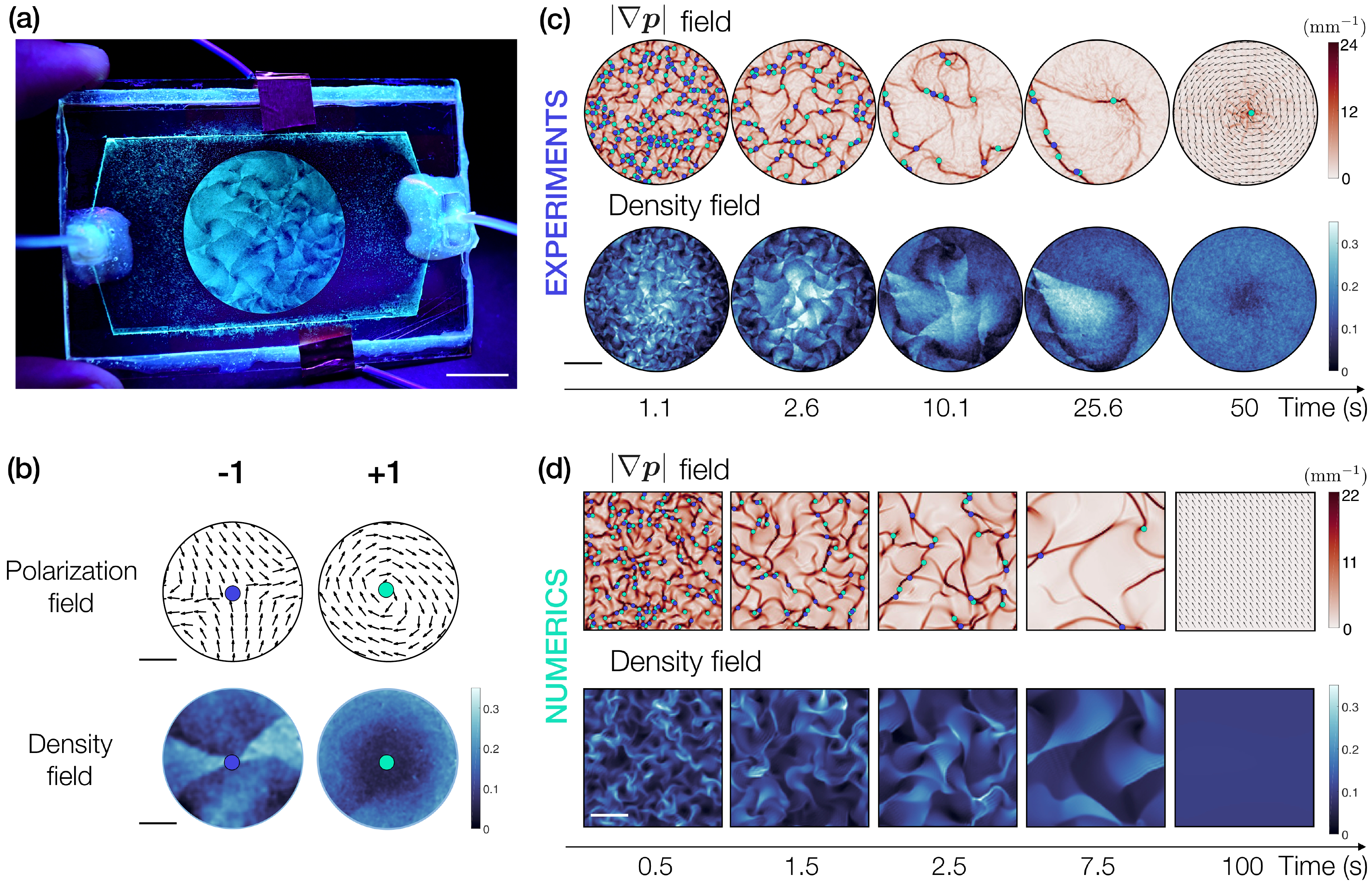}
		\caption{{\bf Ordering dynamics of flocking fluids: experiments and simulations.} 
		\textbf{(a)} Picture of an experiment showing $\sim 4$ millions of colloidal rollers self-assembled into an active polar fluid. The picture reveals typical bow-tie patterns prior to the formation of a pristine vortex pattern at system spanning scale, see Supplementary Movie 1. Scale bar: $1.5\,\rm cm $.
		\textbf{(b)} At the onset of flocking, the flow field  of the active colloidal fluid is plagued by a collection of $\pm1$ topological defects. The density bow ties are centered on $-1$ defects. $+1$ defects are located at the minima of the local density. The color indicates the local packing fraction. Scale bar: $250\,\rm \mu m $. 
		\textbf{(c)} Series of five experimental snapshots of the density field and a map of the local stain magnitude $|\nabla\bm{p}|$. The polarization field is segmented by a fully connected network of domain walls focusing the flow distortions. Each wall coincide with a bow-tie edge in the density field. Scale bar: $2\,\rm mm $.
		 \textbf{(d)} Same representation as in \textbf{(c)} for the numerical resolution of Toner-Tu hydrodynamics, Eqs.~\eqref{eq:toner_tu} with periodic boundary conditions, starting from a homogeneous packing fraction $\rho_0=0.1$ and a random velocity field. ($\lambda=0.7$, $\sigma=5\,\rm mm^{2}.s^{-2}$, $D=10^{-2}\,\rm mm^{2}.s^{-1}$, $\alpha_{0}=100\,\rm s^{-1}$, $\beta=10\,\rm mm^{-2}.s$). See Appendix~\ref{Numerics}. Scale bar: $1.5\,\rm mm $. 	}
		\label{Fig1}
		\end{center}
\end{figure*}

\section{Colloidal flocks}
 To study two-dimensional flocks, we use colloidal rollers which we observe in  microfluidic devices illustrated in Fig.~\ref{Fig1}(a) and Supplementary Movie 1.
 The experimental methods are detailed in Appendix~\ref{Methods} and Refs.~\cite{Bricard2013,Geyer2018}. 
 In short, we start the experiments by filling  circular microfluidic chambers of diameter comprised between $2R=3.5\,\rm mm$ 
 and $2R=3\,\rm cm$ with inanimate polystyrene (PS) spheres of radius $a=2.4\,\rm \mu m$ in  a solution of Dioctyl sulfosuccinate sodium salt (AOT) in hexadecane oil. We let the colloid sediment on the bottom electrode and adjust their area fraction to 10\%. We then take advantage of the Quincke mechanism~\cite{Quincke,Lavrentovich2016} to power the rotation of the PS spheres by applying a DC electric field across two transparent electrodes. Supplementary Movie 1 shows that the microscopic rollers  undergo a flocking transition and collectively organize their  flow into a unique system spanning vortex.

\section{\label{sec:II}Coarsening patterns: bow-ties and domain walls}

\subsection{Self-organization of a colloidal flocking liquid}
To understand how colloidal rollers initially propelling  along random direction self-organize to achieve nearly pristine polar order and steady flows, we measure the instantaneous density and velocity fields, ${\rho=}\rho(\bm r,t)$ and ${\bm{v}=}\bm v(\bm r,t)$, as detailed in Appendix~\ref{Methods}. 
We find that the spontaneous flows are first plagued by a very high density of $\pm 1$ point defects which we detect from the local winding of the polarization field:
{$\bm {p}=\bm{v}/|\bm{v}|$}, see Fig.~\ref{Fig1}(b) and Supplemental Material.
Remarkably, Fig.~\ref{Fig1}(c) and Supplementary Movie 2, showing the magnitude of the instantaneous strain field $|\nabla\bm{p}|$, both reveal that the point defects are not the sole singularities of the velocity field. In fact, they  live on a fully connected  network formed by persistent domain walls separating areas of incompatible orientations. 

The resulting highly tortuous flows are coupled to strong density fluctuations repeating a very characteristic bow-tie pattern, giving the visual impression of a lively folded structure, even though the dynamics is strictly two dimensional, see Fig.~\ref{Fig1}(a),~\ref{Fig1}(c) and Supplementary Movie 3. 
This characteristic motif, emerging from a initially uniform  distribution of rollers, is delimited by two discontinuity lines in the  $\rho$ field  which coincide with  polarization walls crossing at the center of $-1$ defects, see Figs.~\ref{Fig1}(b) and~\ref{Fig1}(c). 
Therefore, as the defects annihilate,  the number of bow ties decreases over time while their typical extent increases,  until they span the whole chamber and vanish.  
We are then left with a heterogeneous polar fluid characterized by a  stable radial density gradient, and a nearly perfect azimuthal flow around a single $+1$ defect consistently reported in flocks of motile grains~\cite{Deseigne2012,Sood2014}, active biofilaments~\cite{Bausch2010} and active colloids~\cite{Bricard2015} in circular geometries. 

\subsection{Coarsening  of Toner-Tu liquids}
The coarsening dynamics illustrated in Fig.~\ref{Fig1} is universal, and does not rely on any  feature specific to the Quincke rollers. 
To demonstrate this universality, we numerically solve the Toner-Tu equations providing a hydrodynamic description of flocking fluid~\cite{Toner95,Toner_review}. They combine mass conservation and the slow dynamics of the velocity field associated to the broken rotational symmetry of the particle orientation. In its minimal form, Toner-Tu hydrodynamics reduces to:
\begin{subequations}\label{eq:toner_tu}
\begin{gather}
\partial_{t}\rho+\nabla\cdot(\rho\bm{v}) = 0\;,\\
\partial_{t}\bm{v}+\lambda\bm{v}\cdot\nabla\bm{v} = (\alpha-\beta v^{2})\bm{v} + D\nabla^{2}\bm{v} -
\sigma\nabla\rho
\end{gather}
\end{subequations}
Eqs.~\eqref{eq:toner_tu} make the intimate relation between the density and orientational order parameter dynamics very clear. From a condensed matter perspective, Eq.~(\ref{eq:toner_tu}b) can be  thought as a Ginzburg-Landau theory for a broken $U(1)$ symmetry, embodied in the vector order parameter $\bm{v}$,  complemented by  self-convection, i.e. $\lambda\bm{v}\cdot\nabla\bm{v}$, and an ordering field $\sigma\nabla\rho$. 
Furthermore, mass conservation naturally couples 
velocity and density fluctuations. From a fluid mechanics perspective Eqs.~\eqref{eq:toner_tu} are akin to the hydrodynamics of a compressible Newtonian fluid 
of kinematic viscosity $D$, 
spontaneously flowing at average speed $v=\sqrt{\alpha/\beta}$ in response to the active drag force $(\alpha-\beta v^{2})\bm{v}$.
Whereas all the coefficients in Eqs.~\eqref{eq:toner_tu} can in principle be density-dependent~\cite{Geyer2018}, for simplicity, we henceforth treat them as constants, with the exception of $\alpha=\alpha_{0}(\rho-\rho_{\rm c})
$ and $\beta=\beta_{0}\rho$
, where $\rho_{\rm c}$ is the critical density of the flocking transition, and $\alpha_{0}$, $\beta_{0}$ are two constants. This choice
guarantees a transition towards collective motion upon increasing the particle density above $\rho_{\rm c}$, as well as the saturation of the flow speed towards the speed $v_{0}=\sqrt{\alpha_{0}/\beta_{0}}$ of individual Quincke rollers when $\rho\gg\rho_{\rm c}$,  deep in the ordered phase.

Using periodic boundary conditions and initializing our numerical resolution of Eqs.~\ref{eq:toner_tu} with homogeneous density and random velocity, we find a coarsening dynamics strikingly similar to our experiments, see Fig.~\ref{Fig1}(d) and Supplementary Movies 4 and 5.
 The numerical methods are detailed in Appendix~\ref{Numerics}. The same bow-tie patterns and domain-wall networks are consistently observed over a range of hydrodynamic parameters as detailed in Supplemental Material. 

Two comments are in order. Our observations in flocking fluids are in stark contrast with the patterns found when solving the (passive) Ginzburg--Landau equation corresponding to $\lambda=\sigma=0$ in Eqs.~\eqref{eq:toner_tu}, and  describing for instance the relaxational dynamics of passive $XY$ ferromagnets (see e.g. Refs.~\cite{Huse93,Bray2002} and Supplementary Movie 6). 
For a passive Ginzburg--Landau dynamics, no mechanism can stabilize domain walls when a $U(1)$ symmetry is spontaneously broken. Hence, all domains walls possibly formed upon a rapid quench vanish diffusively leaving all orientational singularities at the core of $\pm 1$ pointwise vortices, see also Supplemental Material. Another noticeable difference between the defect structure of active and passive polar fluids is that $+1$ vortices in the shape of asters are present only at the very early stage of the dynamics in polar flocks, while they prevail over the full dynamics in passive Ginzburg--Landau systems.  

\begin{figure*}[t!]
	\begin{center}
	\includegraphics[width=1\textwidth]{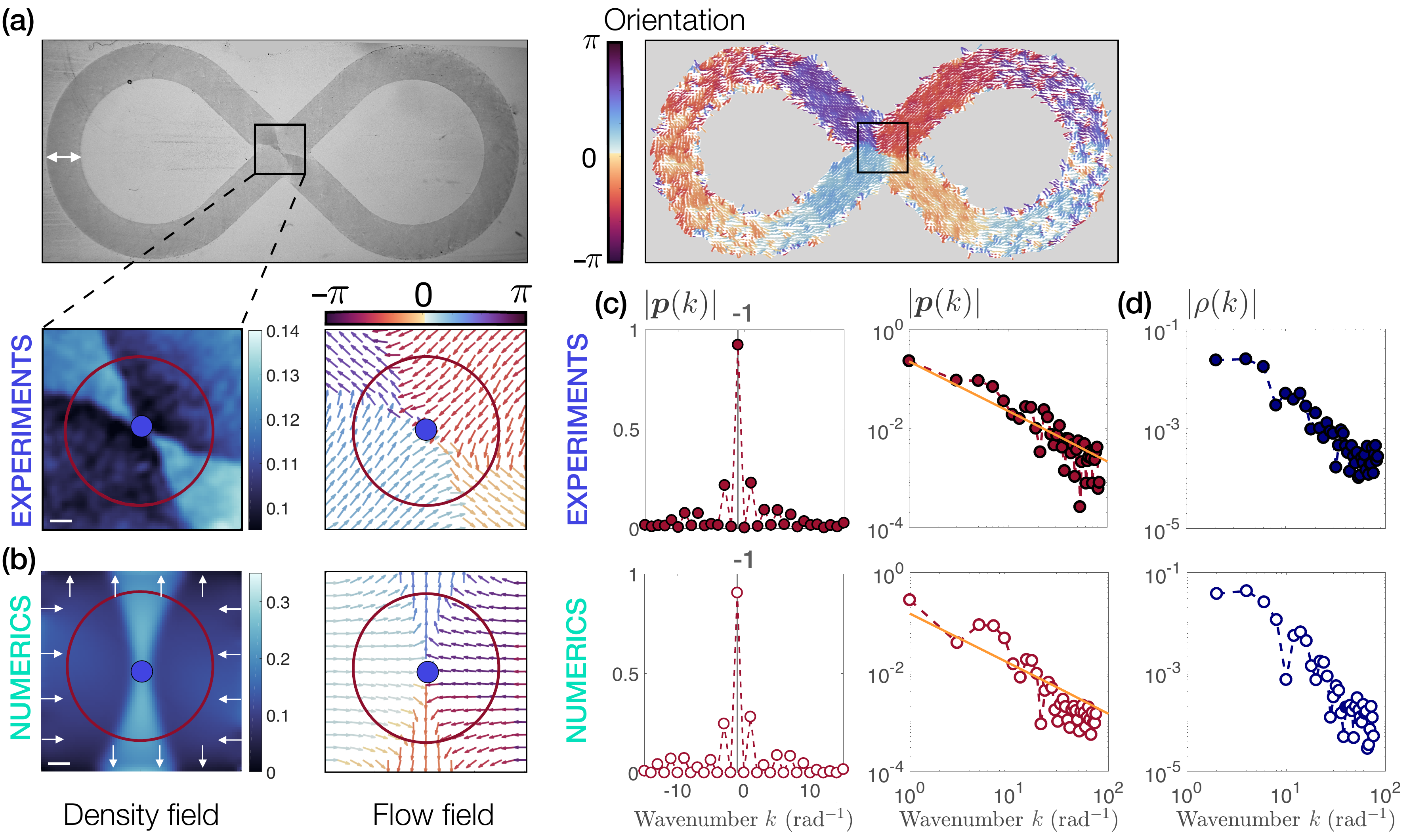}
		\caption{{\bf $-1$ topological charges are dressed with four domain walls and a density bow tie.} \textbf{(a)} Top: Picture of a polar liquid flowing in microfluidic channel in the shape of the figure of eight and corresponding polarization field. The polarization vectors  are colored according to their orientation reveal an isolated $-1$ defect. 
		Channel width: $\rm 5\,mm$. 
		Bottom: Polarization  and density fields  in the vicinity of the $-1$ defect center. The bow-tie structure is clearly visible in the absence of other defects. A single $-1$ charge is stabilized by the imposed direction of the flow field at the boundaries. The density is represented by the local packing fraction. {Scale bar: $1\,\rm  mm$.}
		\textbf{(b)} Numerical simulation of an isolated -1 defect. Density and polarization fields. Boundary conditions described in Appendix~\ref{sec:Numerics}. Same colormaps as in \textbf{(a)}. Scale bar: $150\,\rm \mu m$.
		\textbf{(c)} Angular power spectra of the polarization field computed along the circle showed in \textbf{(a)}. The polarization field is not merely reduced to a single angular mode as in ideal anti-vortex with hyperbolic streamlines. Top: Experimental data. Bottom: Numerical resolution of the Toner--Tu equations Eqs.~\eqref{eq:toner_tu} in the geometry showed in \textbf{(b)}.
		The solid line corresponds to the power law of the theoretical $-1$ defect geometry sketched in Fig.~\ref{Fig:shamrock}(a): four wedge-shaped regions where  $\rho$ and $v$ are piecewise constant are separated by four domain walls emanating from the defect center. 
		\textbf{(d)} Angular power spectra of the density field computed along the circle in \textbf{(a)}. Same power spectra and same observation as in \textbf{(c)}. Both $|\bm p(k)|$ and $|\rho(k)|$ decay algebraically revealing the singular nature of the bow tie patterns. 
		}
		\label{Fig:minus_one}
		\end{center}
\end{figure*}
\section{Morphology of the topological defects in flocking active matter}
In this section we elucidate the atypical geometry of the topological defects of flocking matter and explain how the $-1$ charges shape the emergent domain-wall network and bow-tie patterns discussed in Section \ref{sec:II}.

\subsection{+1 Vortices}
\label{sec:vortex}
It is useful to recall first what determines the morphology of the $+1$ topological defects, which is clearly illustrated by the final vortex pattern of Fig.~\ref{Fig1}(c), see also Ref.~\cite{Bricard2015}. Mass conservation, Eq.~(\ref{eq:toner_tu}a), favors exclusively the emergence of $+1$ vortices, as opposed to asters and spirals~\cite{Kruse2004}. They  correspond to divergenceless azimuthal flows dressed with a radial density gradient extending over system spanning scales. We can understand this pattern from the stationary and long wave-length limit of Eqs. \eqref{eq:toner_tu}, where Eq. (\ref{eq:toner_tu}b) reduces to the balance between convection and density gradients, i.e.
\begin{equation}\label{eq:force_balance}
\lambda\bm{v}\cdot\nabla\bm{v} + \sigma\nabla\rho = \bm{0}\;.
\end{equation}
 We provide a detailed solution of Eq.~\eqref{eq:force_balance} in Appendix~\ref{sec:vortex_appendix}. But we can readily gain some insight on the $+1$ vortex geometry by   neglecting the spatial variations of the flow speed away from the defect center. This assumption is consistent with our experiments where the flow speed hardly depends on the local density sufficiently far from the defect center, see Supplemental Material. As they bend the streamlines, $+1$ defects induce a transverse centrifugal acceleration  $v_0^2/r$ at a distance $r$ from the vortex center, which is balanced  by a radial density gradient over the same scale: $\sigma\partial_r\rho = \lambda v_0^2/r$.
 
 This competition explains why $+1$ defects are located at the minima of the $\rho$ field in Figs.~\ref{Fig1}(a) and~\ref{Fig1}(b). Moreover, it also  reveals that, unlike in liquid crystals, there is no intrinsic length scale setting the size of the defect core. Both the density and flow gradients around a $+1$ vortex are  determined by the system geometry  as well as the position of other defects~\cite{Bricard2015}.
\subsection{\label{sec:bow_tie}Anti-vortices, bow ties and domain walls}
\begin{figure*} 
	\begin{center}
	\includegraphics[width=1\textwidth]{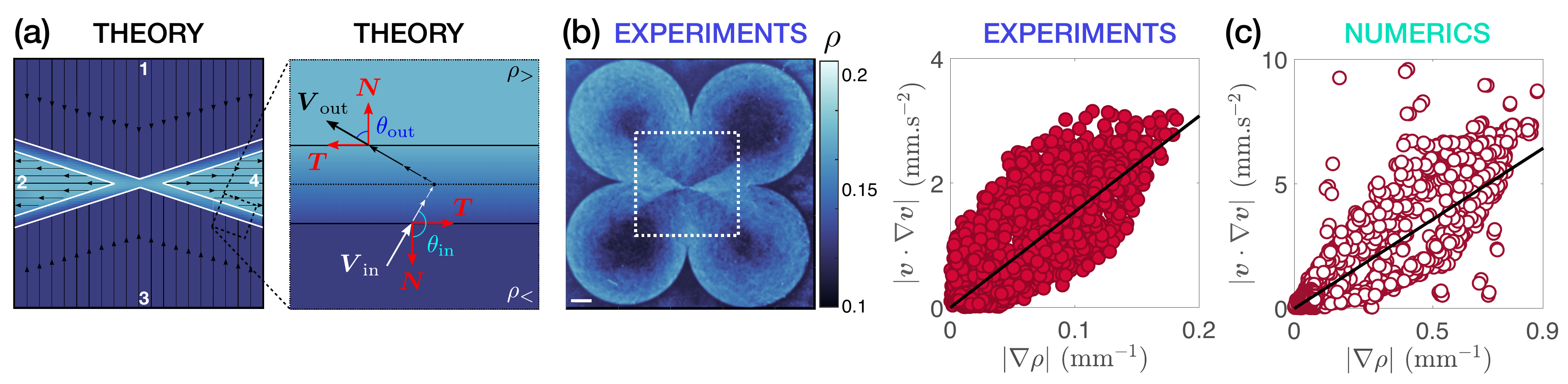}
		\caption{{\bf Bow ties and domain walls emerge from the competition between self-convection and pressure.} 
	\textbf{(a)}. Left: Sketch of the ideal $-1$ defect configuration dressed with four domain walls. The density and flow speed are both piece-wise uniform. Right: close-up on the domain wall and definition of the local tangent $\bm T$ and normal $\bm N$ unit vectors at the domain wall separating the acute and the obtuse regions. 
	\textbf{(b)}. Left: Density field in a shamrock chamber. The density is represented by the local packing fraction. Scale bar: $1\,\rm  mm$. Right: Measuring the stationary flow and density fields in the square region delimited in \textbf{(b)} left, we plot the magnitude of the convective term $|\bm{v}\cdot\nabla\bm{v}|$ against the density gradient $|\nabla \rho|$. The slope of the solid line is given by the ratio  $\sigma/\lambda$ measured independently by fitting the density profiles around the $+1$ defects hosted by  the circular petals, see Supplementary Information. The collapse of all data points demonstrates that density gradients locally balance the convective pressure to stabilize the singular bow-tie patterns. 
		\textbf{(c)}. Same measurements as in \textbf{(b)} for the numerical solution of Eqs.~\eqref{eq:toner_tu} in the geometry of Fig.~\ref{Fig:minus_one}(b). The straight line has a slope {$\sigma/\lambda=7.1\,\rm mm^{2}\;s^{-1}$, see Appendix~ \ref{sec:Numerics}}.}
		\label{Fig:shamrock}
		\end{center}
\end{figure*}

In contrast to the vortex flow discussed in the previous section, there exists no available characterization, or theory, for the distortions induced by a $-1$ topological charge in flocking matter.  
To gain insight about the structure of anti-vortices,
we use a microfluidic chamber having the shape of 
the figure of eight showed in Fig.~\ref{Fig:minus_one}(a), such as to guarantee the existence of an isolated $-1$ topological charge in the polarization field. 
As prominent in Fig.~\ref{Fig:minus_one}(a), the fluid self-organizes in 
a ``bow-tie'' motif characterized by four wedge-shaped regions, where the density and velocity are spatially uniform, separated by sharp boundaries across which both fields changes discontinuously. 
These qualitative observations are quantitatively confirmed by the angular spectra showed in Figs.~\ref{Fig:minus_one}(b) and~\ref{Fig:minus_one}(c). 
They correspond respectively to experiments and numerical solutions of Eqs.~\eqref{eq:toner_tu} in the geometries showed in Fig.~\ref{Fig:minus_one}(a). 
In both cases,  we measure the polarization field $\bm{p}$ at a distance $r$ from the vortex core and Fourier transform it with respect to the azimuthal angle $\phi$: $\bm{p}(r,\phi)=\sum_k \bm{p}_{k}(r)\exp(2\pi i k)$.
Unlike in passive systems where the angular spectra would be solely captured by the $k=-1$ Fourier component,
Fig.~\ref{Fig:minus_one}(b) reveals that a {$-1$} defect excites all polarization modes in a flocking liquid. 
Furthermore, the algebraic decay of the two power spectra measured in experiments and simulations confirms the singular nature of the flow field. The $|\bm p(k)|\sim k^{-1}$ scaling, which corresponds to the Fourier transform of a piecewise constant function at large $k$ values, confirms the formation of four genuine domain walls emanating from the defect center and separating four uniform regions of incompatible orientations, see Figs.~\ref{Fig:shamrock}(a) and \ref{Fig:minus_one}(b).  
The same features are observed in the azimuthal spectra of the density field, Fig.~\ref{Fig:minus_one}(c). The bow tie pattern is delimited by four density discontinuities 
along the four domain walls of the polarization field. We therefore conclude that the domain-wall network   guiding the coarsening dynamics showed in Fig.~\ref{Fig1} emerges from the extended singularities dressing the  $-1$ topological charges of flocking matter. 

\subsection{Focusing the strain field and density gradients along stationary domain walls.}
Our experiments and simulations demonstrate that flocking fluids do not feature perfect anti-vortices, but can support long-lived domain walls focusing all orientation and density gradients. This observation challenges our intuition based on broken $U(1)$ phases in equilibrium and begs for a quantitative explanation. 

\subsubsection{Flocking matter cannot host ideal anti-vortices}
Let us first consider a hypothetical perfect anti-vortex having constant flow speed, $v={\rm const}$, and orientation $\bm{p}=\cos\phi\,\bm{e}_{x}-\sin\phi\,\bm{e}_{y}$, representing the far-field configuration around a classical $-1$ defect located at the origin.  
The associated streamlines are hyperbolas, described by the implicit equation $xy=r_{0}^{2}/2$, with $r_{0}$ the minimal distance of the streamline from the origin attained when $\phi=(\pi/4)\,n$, with $n=\pm 1,\pm 3$. 
In steady state, noting  $\partial_{\parallel}=\bm{p}\cdot\nabla$ differentiation along a streamline, we can recast mass conservation, Eq. (\ref{eq:toner_tu}a), into
\begin{equation}\label{eq:stream_line_continuity}
\frac{\partial_{\parallel}\rho}{\rho} = \frac{\cos 2\phi}{r}\;.
\end{equation}
Defining the curvilinear coordinate $s$, and setting $s=0$ at $(r,\phi)=(r_{0},\pi/4)$, we can then readily integrate Eq.~\eqref{eq:stream_line_continuity} along a streamline to express the density variations as
\begin{equation}\label{eq:stream_line_density}
\log\left[\frac{\rho(s)}{\rho(0)}\right] =  \int_{0}^{s}{\rm d}s'\,\frac{\sqrt{r^{4}-r_{0}^{4}}}{r^{3}}\;.
\end{equation}
Eq.~\eqref{eq:stream_line_density} implies that the fluid density$\rho$ monotonically decreases in the upstream direction ($s<0$). However as as $v^{2}=v_{0}^{2}(1-\rho_{\rm c}/\rho)$, the flow speed decreases with $\rho$, and therefore Eq.~\eqref{eq:stream_line_density}  contradicts the initial assumption of a uniform flow speed. In fact, Eq.~\eqref{eq:stream_line_density}  implies that  flocking motion would be suppressed at a sufficiently large distance from the defect center as $v$ would vanish once $\rho<\rho_{\rm c}$.  We therefore conclude that  macroscopic polar flocks cannot host perfect anti-vortices. The $-1$ defects of polar active matter are inherently associated to inhomogeneity in the density and flow speed. We 
 show below how they can nevertheless be partially relieved by focusing orientational variations along one dimensional singularity lines  emanating from the defect center. 
\begin{figure*} 
	\begin{center}
	\includegraphics[width=1\textwidth]{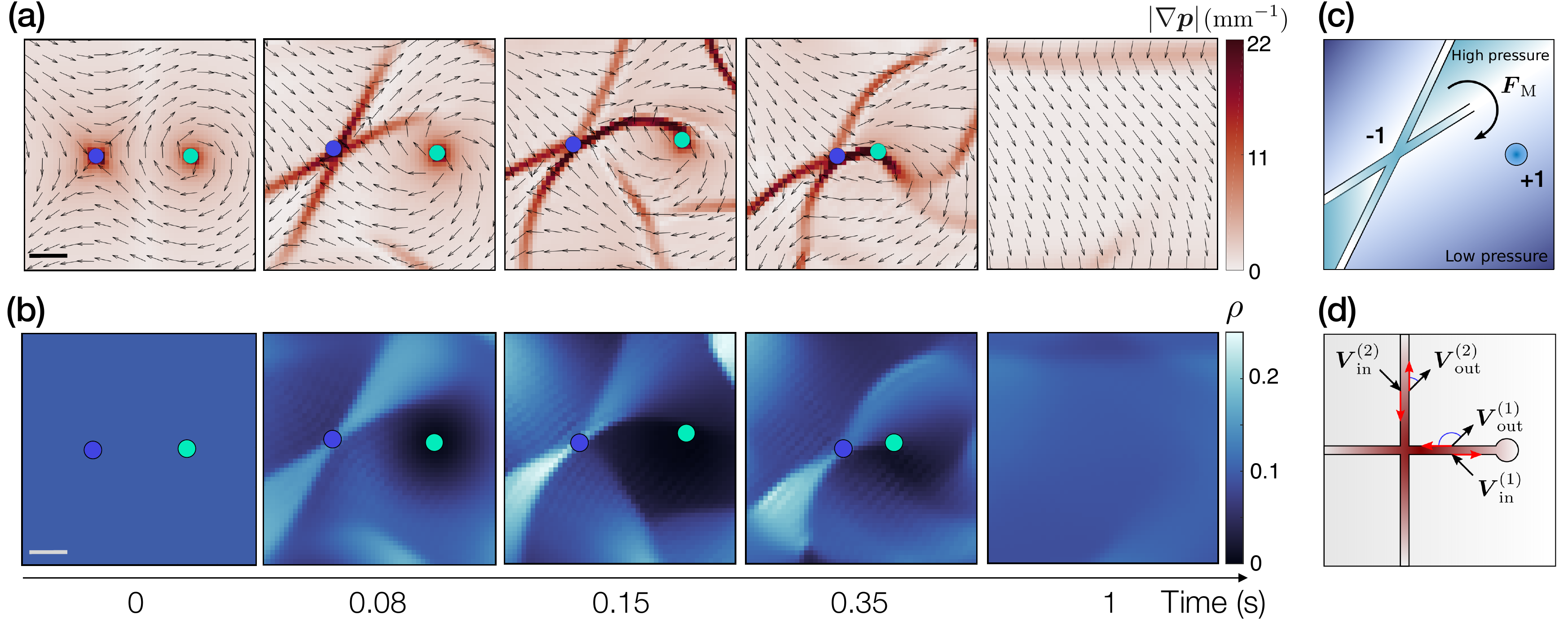}
		\caption{{\bf Self-advection and pressure gradients link topological charges with polarization walls.} 
		\textbf{(a)}. Solving Eqs.~\eqref{eq:toner_tu} we study the dynamics of the polarization and density field around a pair of topological defects in a periodic box. At $t=0$ the flow field is the superposition of the ideal vortex and anti-vortex (hyperbolic stream lines). The flow gradient first focus to form sharp domain walls growing from the center of the $-1$ defect. The domain walls subsequently bend and rotate to link the $+1$ charge. Once the defects are linked, the $+1$ charge approaches and annihilates with the $-1$ defect, thereby yielding a perfect uniaxial order. Scale bar: $500\,\rm \mu m $.
		\textbf{(b)}. Corresponding evolution of the initially uniform density field around the defect pair. A bow-tie pattern emerges as the domain walls grow and eventually vanishes once the two topological charges annihilate. Scale bar: $500\,\rm \mu m $.
		\textbf{(c)}. Illustration of the pressure-driven dynamics of the domain walls around a $-1$ defect. 		%
		\textbf{(d)}. Once the four domain walls form four $\pi/2$ wedges, and link $+1$ charge, their opening remains fixed provided that the incoming and outgoing flows satisfy the horizontal mirror symmetry.  
		}
		\label{Fig:branching}
		\end{center}
\end{figure*}

\subsubsection{Strain focusing around $-1$  topological charges}
To gain further physical insight on the strain focusing around $-1$ defects, we perform an additional set of experiments in the shamrock geometry showed in Fig.~\ref{Fig:shamrock}(b). 
The resulting flow field hosts a $-1$ defect at the shamrock's center and one $+1$ defect in each leave. We can therefore measure the ratio {$\sigma/\lambda$} by fitting the density profiles in the four vortices, as detailed in Supplementary Information. 
We then plot the magnitude of the convective acceleration $|\bm{v}\cdot\nabla\bm{v}|$ as a function of the modulus of the local density gradient measured in the vicinity of the defect center. 
Doing the same analysis for the numerical flow field of  Fig.~\ref{Fig:minus_one}(b), we find that all experimental and numerical data measured around the $-1$ charge consistently obey the scaling relation $|\bm{v}\cdot\nabla\bm{v}|=(\sigma/\lambda)|\nabla\rho|$, Figs.~\ref{Fig:shamrock}(b) and~\ref{Fig:shamrock}(c). 
This relationship establishes that the polarization walls and density bow ties centered on $-1$ defects emerge from the competition between self-advection and density gradients generic to all  spontaneously flowing liquids.  

Informed by our experimental observations, we consider a piece-wise uniform configuration for $\rho$ and $\bm p$ in four wedge-shaped regions as defined in Fig.~\ref{Fig:shamrock}(a). 
 To find a solution of the full active-hydrodynamic problem, we must now show that the density and flow speed can interpolate between  two adjacent bulk values within finite domain walls, while the orientation $\bm{p}$  rotates, either continuously or discontinuously, by $\pi/2$. 
 Switching again to streamline coordinates we write $\nabla=\bm{p}\,\partial_{\parallel}+\bm{p}^{\perp}\partial_{\perp}$, with $\bm{p}^{\perp}$ the counter-clockwise normal to $\bm{p}$, and Eq.~ \eqref{eq:force_balance} takes the form
\begin{equation}\label{eq:domain_wall}
\bm{p}\,\partial_{\parallel}\left(\frac{1}{2}\,\lambda v^{2}+{\sigma\rho}\right)+\bm{p}^{\perp}\left({\sigma\partial_{\perp}\rho}+\kappa v^{2}\right) = \bm{0}\;,
\end{equation}
where $\kappa$ is the signed curvature of the streamline. Similarly to the case of $+1$ vortices (see Sec.~\ref{sec:vortex}), a transverse density gradient,  $\partial_{\perp}\rho$, is necessary to counterbalance the centrifugal  acceleration $\kappa v^{2}$  resulting from the bending of the streamlines. 
However, because of the microscopic thickness of domain walls, which sets the radius of curvature of the streamlines, 
such a density gradient would have to be significantly large at any finite speed $v$, thereby making the flow potentially unstable against density fluctuations. 
Alternatively, the fluid can continue traveling in a straight trajectory (i.e. $\kappa=0$) until reaching the domain wall centerline. 
The infinite curvature originating from the abrupt $\pi/2$ rotation is then compensated 
by a vanishing flow speed without requiring any  transverse gradient  ($\partial_{\perp}\rho=0$).
In this case, the second term on the left-hand side of Eq.~\eqref{eq:domain_wall} vanishes, whereas the balance between self-advection and pressure gradients along the longitudinal directions results in the conservation law \begin{equation}\label{eq:bernoulli}
\frac{1}{2}\,\lambda v^{2} + \sigma\rho = {\rm const}\;.
\end{equation}
This relation is a flocking-matter analogue of the Bernoulli law of inviscid flows. Having assumed that, deep in the ordered phase, alignment interactions ensure the local relation $v^2=v_{0}^{2}(1-\rho_{\rm c}/\rho)$, and using Eq.~ \eqref{eq:bernoulli} to solve Eq.~ (\ref{eq:toner_tu}b) with respect to $v$, finally gives the an approximated solution for the configuration of the density and flow speed within domains walls. Calling $\rho_{<}$ and $\rho_{>}$, with $\rho_{<}\le\rho_{>}$ and $\rho_{0}=(\rho_{<}+\rho_{>})/2$, the density of either one of the two regions separated by the same domain wall [Fig.~ \ref{Fig:shamrock}(a)], we find
\begin{subequations}\label{eq:boundary_layer_solution}
\begin{gather}
\rho(s) = \rho_{\lessgtr} + \frac{\lambda}{2\sigma}\,\left[v_{0}^{2}-v^{2}(s)\right]\;,\\
v(s) = v_{0} \tanh\left(\frac{s}{\xi_{\lessgtr}}\right)\;,	
\end{gather}	
\end{subequations}
where we approximated $v\approx v_{0}$ away from the domain wall under the assumption that $\rho_{\lessgtr}\gg\rho_{\rm c}$.  The  thickness of the domain wall is given by $\xi_{\lessgtr} = \sqrt{2D/(\alpha_{0}\rho_{\lessgtr})}$, as detailed in Appendix~\ref{sec:boundary_layer_appendix}. 
This typical domain-wall solution  allows the flow and density field to interpolate between seemingly incompatible domains. It further confirms that self-advection, aligning interactions and pressure gradients are the basic ingredients dressing the  $-1$ topological defects with orientational domain walls and density bow ties.  

However, we still need explaining the magnitude of the orientational and density discontinuities across a domain wall. Integrating the mass conservation relation in the rectangular region of arbitrary size showed in Fig.~\ref{Fig:shamrock}(a), relate the density jump to the orientation mismatch between the regions across a domain wall sketched in Figs.~\ref{Fig:shamrock}(b) {and~\ref{Fig:shamrock}(c)}. As detailed in Appendix~\ref{sec:boundary_layer_appendix}, taking $\rho_{>}=2\varphi\rho_{0}$ and $\rho_{<}=2(1-\varphi)\rho_{0}$, with $1/2\le\varphi\le 1$, deep in the flocking phase this relation reduces to   
\begin{equation}\label{eq:density_fraction}
\varphi = \frac{\cot\theta_{\rm in}}{\cot \theta_{\rm in}-1} = \frac{\tan \theta_{\rm out}}{\tan\theta_{\rm out}+1}\;,
\end{equation}
 and where $\theta_{\rm in}$ and $\theta_{\rm out}$ are the angles between the incoming and outgoing streamlines and domain wall and are such that $|\theta_{\rm in}|-|\theta_{\rm out}|=\pi/2$, see Fig.~\ref{Fig:shamrock}(a). 

Toner--Tu hydrodynamics, however, cannot  prescribe the absolute magnitude of the orientational discontinuity across the polarization walls, the orientation and the opening of the bow-tie pattern. Analogously to the core radius $a$ of $+1$ vortices,  these three geometrical features  are determined by the far-field configuration of the velocity field $\bm{v}$. They are therefore set by the boundary conditions, and interactions with the other defects populating the system, which we discuss in the next section.

\section{Emergent domain-wall networks and defect interactions}

We now need to explain how the polarization domain walls form the prominent network linking all $\pm 1$ defects in polar liquids when actively organizing their flows, see Figs.~\ref{Fig1},~\ref{Fig:Coarsening} and Supplementary Movies 2 and 4.

\subsection{Interactions between domain walls and topological charges}
To illuminate this spectacular example of topology-driven self-organization, we simulated the annihilation dynamics of a single pair of $\pm 1$ defects, whose initial configuration consists of a perfect vortex-anti-vortex pair in a homogeneous Toner--Tu fluid. 
The image sequence of Fig.~\ref{Fig:branching} illustrates this three-step dynamics. 
 Shortly after the beginning of the simulation, the $-1$ defect evolves towards the typical bow-tie structure, distinguishing two acute and two obtuse wedge-shaped regions separated by domain walls, where the velocity field rotates discontinuously by $\pi/2$. 
Although the neighboring $+1$ defect is initially disconnected from the four singular lines, the East-West-oriented wall subsequently bends and rotates to irreversibly connect to the $+1$ vortex center. 
Remarkably, it is only once the link is formed, that  the $\pm 1$ charges approach one another along the  polarization wall to eventually annihilate and yield a pristine uniaxial flow. 

\begin{figure*}
	\begin{center}
	\includegraphics[width=1\textwidth]{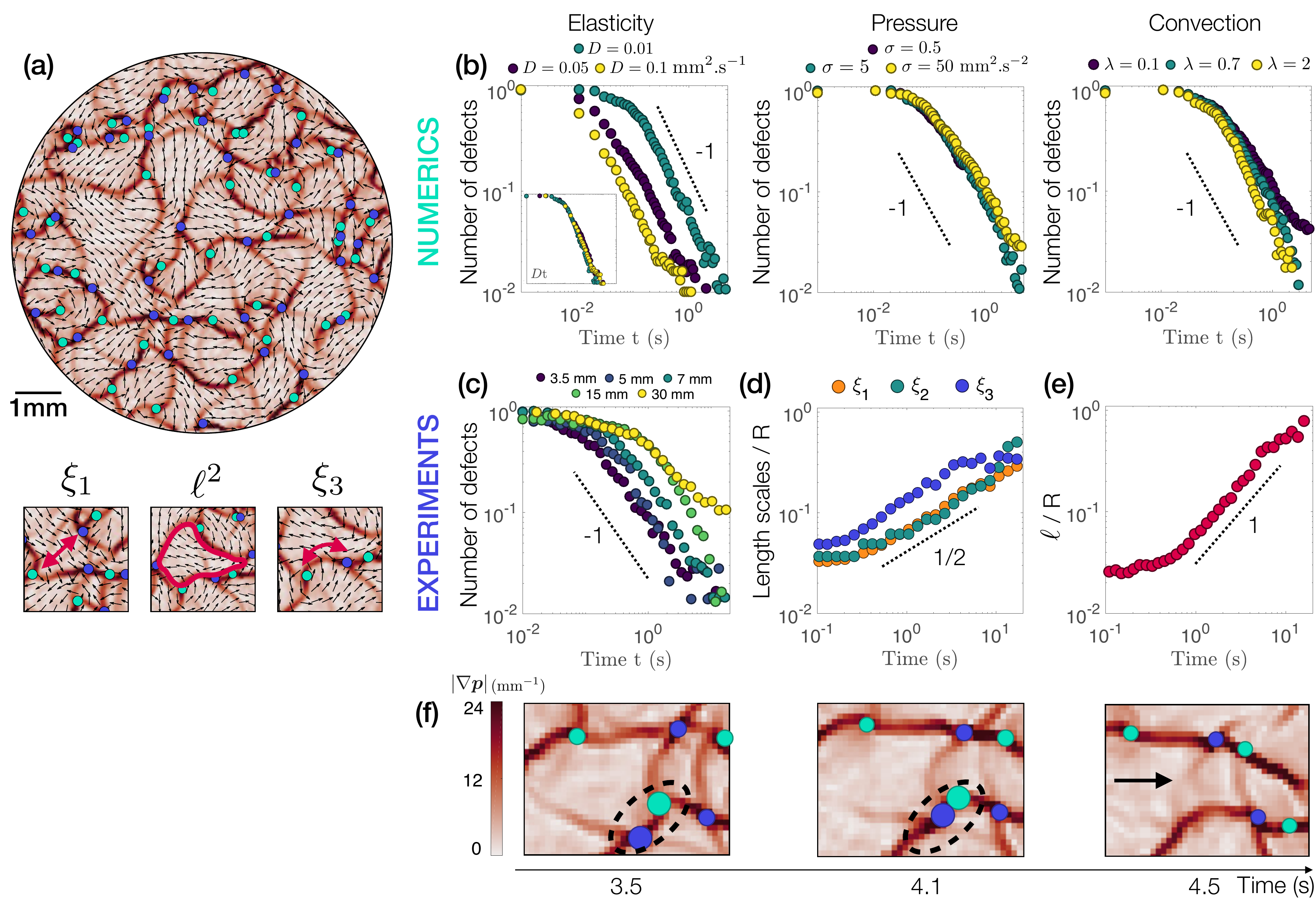}
	\caption{{\bf Orientational elasticity rules a self-similar phase ordering kinetics.}
	 \textbf{(a)}. Snapshot of the polarization field (experiments). The red lines indicate the domain walls, the light green (resp. dark blue) circles indicate the position of the $+1$ (resp. $-1$) topological charges. The surrounding regions form  domains of incompatible uniform polarization. %
	 \textbf{(b)}. Time evolution of the total number of topological defects. We take advantage of the numerical resolution of Eqs.~\eqref{eq:toner_tu} to vary independently the three essential hydrodynamic parameters $D$, $\sigma$ and $\lambda$, setting the magnitude of the orientational elasticity, of  the compressibility and of the self-advection coefficient of the flocking liquid. 
	 All the constant parameters are the same as in Fig. 1.  The number of the defects plaguing the active flows decays as $1/t$ for all hydrodynamic parameters. The long-time coarsening dynamics is chiefly determined by the orientational elasticity of the polar liquid and all data collapse on a single master curve when plotted against $Dt$ (left panel inset). 
	 \textbf{(c)}. Experimental characterization of the phase ordering kinetics in chambers of increasing size. In agreement with our simulations the number of defects decays as $1/t$. 
	 \textbf{(d)}. Equivalently the absolute, $\xi_1$, and curvilinear, $\xi_3$, distances between the defects defined in \textbf{(a)} grow algebraically as $t^{1/2}$ and so does the correlation length $\xi_2$ of the $\bm p$ field, see also Supplemental Material. 
	 \textbf{(e)}. The typical size of the polarization domains $\ell$, however, grows ballistically in time.
	 \textbf{(f)}. Three  snapshots of the strain field corresponding to the experiment showed in {\bf (a)}.  When a defect pair annihilates (dashed line) the domain wall emanating from them bend an eventually vanishes. 
	 As a result defects defining the vertices of the polarization network end up living along the domain walls. The typical distance between the defects hence grows slower than the distance between the network vertices $\sim\ell$. 
	}
		\label{Fig:Coarsening}
		\end{center}
\end{figure*}

Until now  our experimental and numerical findings consistently indicate that the spin-wave elasticity associated to the polarization field plays a secondary role in the emergence of the domain wall and bow tie patterns. Ignoring this contribution, $D$ term in Eq.~(\ref{eq:toner_tu}b), the hydrodynamics of our active fluid is essentially that of an inviscid fluid flowing at constant speed.  
We can use this simplified picture to single out the mechanisms underpinning the rotation of the domain walls at the onset of a fully connected network, see Figs.~\ref{Fig1}(c) and \ref{Fig:Coarsening}(a).
To do so, we consider the typical situation involving two defects of opposite charges sketched in Fig.~\ref{Fig:branching}(c). In the presence of a nearby $+1$ vortex the density mismatch between the  acute and obtuse  regions  delimited by the domain walls is further increased. Therefore, the resulting pressure gradient drives a lift force that rotates the domain walls.  
The only possible stationary state is then  given by the symmetric conformation depicted in Fig.~\ref{Fig:branching}(d), where the domain walls emanating  from the same $-1$ defects are orthogonal and oriented at a $\pi/4$ angle with respect to the incoming and outgoing velocity field.  In order to go beyond this basic symmetry argument, and account for the subsequent attraction between the defect centers, we  introduce below a quantitative theory of topological defect interactions in flocking matter.

\subsection{Topological-defect interactions in flocking matter}
Our minimal theory of defect interactions is inspired by the dynamics of quantized vortices in superfluids, see e.g. Ref.~\cite{Sonin87}.  Given a  topological defect of  position $\bm{R}$ {and velocity $\dot{\bm{R}}$},
we show in Appendix~\ref{sec:vortex_appendix} that {convection} and {density gradients}  result in a net active Magnus force 
\begin{equation}\label{eq:Magnus}
\bm{F}_{\rm M} = \Gamma\,\bm{\epsilon}\cdot(\bm{V}_{\rm far}-\dot{\bm{R}})\;,
\end{equation}
where $\bm{V}_{\rm far}$ is the velocity resulting from the far-field configuration of the flow, $\bm{\epsilon}$ the antisymmetric tensor, with $\epsilon_{xy}=-\epsilon_{yx}=1$ and $\epsilon_{xx}=\epsilon_{yy}=0$, and $\Gamma$ an effective drag coefficient associated with the configuration of the velocity field along the core $\mathcal C$, namely 
\begin{equation}\label{eq:gamma}
\Gamma = \lambda \oint_{\mathcal C}{\rm d}{\bm \ell}\cdot(\rho\bm{V})\;,
\end{equation}
where $\bm V=\bm{v}-\bm V_{\rm far}$ is the contribution  to the total velocity  originating from the defect. $\bm F_{\rm M}$ reflects the transverse response of the $\pm 1$ defects when driven by an external flow, and is akin to the conventional Magnus force experienced by vortices in Euler fluids~\cite{Lamb}. 
In the case of an isolated defect, $\bm{F}_{\rm M}$ is either vanishing or balanced by the longitudinal drag force $\bm{F}_{\rm D}=-\zeta\dot{\bm{R}}$, with $\zeta$ a drag coefficient, resulting from the broken Galilean invariance of Eqs.~(\ref{eq:toner_tu}b) and reflecting, in particular, the slow spatiotemporal variations of the order parameter $v$ originating from the motion of the defect core.

In the presence of other defects, however, both forces must balance the classic Coulomb $\bm{F}_{\rm C}$, resulting from the orientational elasticity of the polarization field $\bm{p}$ (see e.g. Ref.~\cite{Chaikin}). Under the simplifying assumption that the three forces can be computed independently, the force balance condition $\bm F_{\rm M}+\bm F_{\rm C}+\bm F_{\rm D}=\bm 0$ is readily recast in an equation of motion for the interacting defect centers $\bm{R}_{i}$, with $i=1,\,2,\,3,\,\ldots$. These equations of motion read:
\begin{equation}\label{eq:defectmotion}
(\zeta_i \mathbb{1}+\Gamma_i\bm \epsilon)\cdot\dot{\bm R}_i= \Gamma_i\bm{\epsilon}\cdot\sum_{j \neq i} \bm V_{j}+2\pi K_{i}\sum_{j \neq i} \Omega_i\Omega_j \frac{\bm R_i-\bm R_j}{|\bm R_i-\bm R_j|^2}\;,
\end{equation}
{where $\mathbb{1}$ is the identity tensor, $K_{i}=\zeta_{i}D$ is the orientational stiffness of the polarization field, $\Omega_{i}=\pm 1$ is the winding number of the vortices and we have approximated $\bm V_{\rm far}=\sum_{j \neq i} \bm V_{j}$}. Although evidently simplified, this theory does not only shed light on the linking of topological defects by polarization walls, but also explain the subsequent annihilation of  defects cruising along the resulting singularity network. Considering again the two defect configuration of Fig.~\ref{Fig:branching} where two defects of opposite charge are located at $\bm R_+$ and $\bm R_-$, Eq.~\eqref{eq:defectmotion} implies that the dynamics in the direction transverse to $\bm R_+$ and $\bm R_-$   stops only when $\Gamma_\pm$ vanishes. In particular $\Gamma_-$ can be calculated from Eq. \eqref{eq:gamma} in the form
\begin{equation}
\label{eq:gamma_minus}
\Gamma_{-} = \lambda\sum_{n=1}^{4}\rho{^{(n)}}\mathcal{L}^{(n)}\left[V_{\rm in}^{(n)}\sin \theta_{\rm in}^{(n)}+V_{\rm out}^{(n)}\sin \theta_{\rm out}^{(n)}\right]\;, 
\end{equation}
where the index $n$ denotes each of the four domain wall comprising the core and $\mathcal{L}^{(n)}$ their length. We find that the right-hand side of Eq. \eqref{eq:gamma_minus} vanishes in the geometry of Fig.~\ref{Fig:branching}(d), when the polarization walls are orthogonal and of equal length, in which case the terms in the sum have equal magnitude and alternating signs. 
Once this situation is reached the Magnus drag remains vanishingly small and the defect dynamics of Eq.~(\ref{eq:defectmotion}) is then purely longitudinal. In other words, a remarkable prediction of our theory is that flocking liquids actively organize their flows to form  polarization walls and density bow ties making the topological defect dynamics virtually indistinguishable from a  passive polar materials devoid of such intricate excitations~\cite{Bray2002}, see Supplementary Movies {2, 4 and 6}.  

\section{Self-Similar phase ordering kinetics}
We are now equipped to quantitatively explain the phase ordering kinetics of flocking matter. 
At long time, when the defect dynamics is purely longitudinal ($\Gamma_i=0$), and restricted to the domain-wall network, Eq.~\eqref{eq:defectmotion}  predicts an algebraic kinetics as in passive $XY$ ferromagnets and polar liquid crystals~\cite{Huse93,Pargellis94,Bray2002,Cugliandolo2011}.  
For instance, as defects annihilate via pair collisions, the number density of $\pm 1$ defect-pairs, $n_{\rm P}$ obeys a first order kinetic equation $\partial_t n_{\rm P}=-k n_{\rm P}$.  
Within this  mean field picture, the time-dependent kinetic constant $k=k(t)$ equates the inverse annihilation time $\tau$. 
The latter can be estimated  by solving Eq.~\eqref{eq:defectmotion} for a pair of $\pm 1$ defects in the limit of vanishing Magnus force, which yields $|\bm R_+-\bm R_-|^2=4\pi\tau D$. This gives $k \sim D/\delta^{2}$, with $\delta$ the typical inter-vortex distance. Finally, taking $\delta \sim 1/\sqrt{n_{\rm P}}$ gives $k \sim Dn_{\rm P}$, from which it  readily follows  the {classic scaling law of defect coarsening of the $XY$ model and related systems: i.e. $\mathcal{N} \sim \int {\rm d}A\,n_{\rm P} \sim (Dt)^{-1}$.  Equivalently, we expect the coarsening dynamics to be self-similar, and to observe a diffusive growth of all structural length scales (up to logarithmic corrections)~\cite{Bray2002}. 
 
 In order to quantitatively confirm that the asymptotic phase ordering of  flocks is  ruled by the elasticity of this active broken symmetry phase, we first perform a series of numerical simulations varying independently the three essential parameters $\lambda$, $\sigma$ and $D$ of Toner-Tu hydrodynamics. 
 Measuring the number of defects $\mathcal{N}$ in Fig.~\ref{Fig:Coarsening}(b) unambiguously shows that  the defect-annihilation kinetics is chiefly controlled by the elastic/Coulomb interactions, see also Supplemental Material. 
 To further ascertain the self-similar nature of the dynamics, and rule out possible finite size effects, we quantify the temporal evolution of the polarization network geometry   in a series of experiments and simulations spanning an order of magnitude in size. 
 We show our experimental findings in  Figs.~\ref{Fig:Coarsening}{(c), (d) and (e)}, and our consistent numerical results in Supplemental Material.  
Fig.~\ref{Fig:Coarsening}(c) confirms that the number of $\pm 1$ topological defects decays algebraically as $\mathcal N\sim t^{-1}$ regardless of the system size. 
Equivalently, the absolute and curvilinear inter-defect distances $\xi_1$ and $\xi_3$,  defined in Fig.~\ref{Fig:Coarsening}(a),  both obey a diffusive scaling law $\xi\sim t^{1/2}$, see Fig.~\ref{Fig:Coarsening}(d). The absolute distance between the defect cores matches the orientational correlation length $\xi_2$ measured from the spatial decay of the two point function {$C_p(r)=\langle \bm{p}(\bm{r},t)\cdot \bm{p}(\bm{0},t)\rangle$}, which also grows diffusively in time as the colloidal flocks self-organize, see Figs.~\ref{Fig:Coarsening}(d) and Supplemental Material.  

However, Figs.~\ref{Fig:Coarsening}(e) indicates that the size $\ell$ of the  domains delimited by the polarization walls features a much faster ballistic dynamics. This observation, does not  contradict our explanation of the active coarsening process, but  provides an additional insight on  the interplay between the domain-wall geometry and the topological charge dynamics.
The faster growth of the domain size indeed originates from the fact  that the $\pm 1$ defects  do not only define the vertices of the network, but can also navigate along its edges. The image sequence of Fig.~\ref{Fig:Coarsening}(f) illustrates this crucial aspect of the dynamics. The $\pm 1$ defects interact solely along the edges of the network and each vertex hosts a defect connected up to four defects of opposite topological charge. But,
 when two defects attract and annihilate the transverse domain walls they were attached to bend
  and connect another opposite charge, or fade away. The defect annihilation therefore leaves $\pm 1$ defects living along an edge of the network, away from a vertex.   This intricate yet consistent dynamics  explains why the typical domain size set by the inter-vertex distance exceeds the typical separation between opposite topological charges living along the edges. 
\section{Conclusion}
Combining experiments, {numerical simulations and analytical work,} we have explained how flocks of self-propelled particles suppress the topological excitations of their flow field and self-organize in one of the most stable ordered phase observed in nature. 
This atypical phase ordering dynamics is ruled by the emergence of domain-wall networks shaped by self-advection and density gradient around $-1$ topological charges. 
This lively structure mirrored by density patterns in the form of bow-tie structures has no counterparts in passive systems. Remarkably, by actively constraining their topological charges to cruise and annihilate along polarization walls, flocking matter achieves a self-similar coarsening kinetics characterized by the same diffusive exponent as in passive ferromagnets, superconductors, or {thin films of} liquid crystals. 
The consistent and quantitative agreement between our experimental measurements and the resolution of a minimal hydrodynamic model of active polar flows further confirm the universality of our findings beyond the specifics of colloidal roller experiments. 

A natural question arises from our comprehensive study: how does collective motion emerge over system spanning scales in higher dimensional systems such as bird flocks, polar tissues or 3D synthetic flocks? How does the interplay between density fluctuations and polar order alter the fundamental topological excitations of high dimensional polar active matter and compressible active liquid crystals?
\section*{Acknowledgements}
{This work is partially supported by ANR WTF, Idex Tore and Tremplin CNRS (A.C., Y.P. and D.B.), by Netherlands Organization for Scientific Research (NWO/OCW), as part of the Vidi scheme ({L.A.H.} and L.G.), and by the European Union via the ERC-CoG grant HexaTissue (L.G.).} 

\appendix
\section{Experimental methods}
\label{Methods}
\subsection{Quincke rollers experiments} 
Our experimental setup corresponds to the one introduced in~\cite{Bricard2013,Geyer2018}.
We disperse polystyrene colloids of radius $2.4\,\rm\mu m$ (Thermo Scientific G0500) in a solution of hexadecane including  $5.5\times10^{-2}\,\rm wt\,\%$ of dioctyl sulfosuccinate sodium salt (AOT). 
We then inject the suspension in microfluidic chambers made of two electrodes spaced by a $25\,\rm \mu m$-thick scotch tape. 
The electrodes are glass slides, coated with indium tin oxide (Solems, ITOSOL30, thickness: $80\,\rm nm$). 
In all our experiments, we let the colloids sediment on the bottom electrode and apply a DC voltage of $130\,\rm V$. 
The resulting electric field triggers the so-called Quincke electro-rotation and causes the colloids to roll at a constant speed $v_{0}=0.8\,\rm mm/s$~\cite{Bricard2013,Taylor69}.
The geometry of the microfluidic device is illustrated in Fig.~\ref{Fig1}(a). 
We confine the Quincke rollers inside circular chambers of diameter comprised between $2R=3.5\,\rm mm$ and $2R=30\,\rm mm$. 
The confining circles are made of a $2\,\rm \mu m$-thick layer of insulating photoresist resin (Microposit S1818) patterned by means of conventional UV-Lithography as explained in~\cite{Morin2017}. 
The patterns are lithographed on the bottom electrode. 
We start the experiments by filling homogeneously the microfluidic chambers, and the average packing fraction $\approx10\%$,  is chosen to be far beyond the flocking transition threshold ($ \approx 0.1\%$).
We use the same roller fraction in all the experiments discussed in the main text.

We image the whole chambers with a Nikon AZ100 microscope using a  magnification comprised between 2X and 6X depending on the chamber size, and  record videos with a Luxima LUX160 camera (Ximea) at a frame rate of $200\,\rm fps$. 
We start measuring the flow field  before the application of the DC field and stop recording only once the active flow has relaxed all its singularities, i.e. once it forms a steady vortex pattern spanning the whole circular chamber. 
Every experiments were repeated several times.

\subsection{Velocity and density  fields} 
In the largest  chambers, we cannot detect the individual position of all the rollers with a sufficient accuracy to rely on Particle Tracking Velocimetry (PTV). Instead, we systematically use Particle Imaging Velocimetry (PIV)  to construct the velocity field $\bm{v}$. 
PIV is performed using the  PIVLab  Matlab package, see \cite{Thielicke2014}. The PIV-box size is $83.2\,\rm \mu m \times 83.2\,\rm \mu m$, with an overlap  of a half PIV box between two adjacent measurements. We systematically checked that our findings are qualitatively insensitive to the specific choice of the PIV parameters. The polarization and strain fields are then readily computed with the same spatial resolution from $\bm{v}$.

In order to measure the density field $\rho$, we use the intensity scale of the 8-bit images. In practice, we first subtract the background image (average intensity over 500 subsequent images) to each original image, we then divide the result by its maximal intensity value. Averaging over boxes of $83.2\,\rm \mu m \times 83.2\,\rm \mu m$, we finally reconstruct the density field with the same resolution as for the velocity field. A direct comparison with measurements performed by locating the position of every single colloids using a higher magnification  confirms the accuracy of our method within an accuracy of 10\%. 

To accurately measure the individual velocity and the packing fraction of the colloids,
we track the position of all the rollers with a sub-pixel accuracy using the algorithms introduced by Lu {\em et al.}~\cite{Lu2007} and by Crocker and Grier~\cite{Grier}. When powered with an electric field of magnitude $130\,\rm V$, all colloids roll at a constant speed:
$
v_{0}= 0.80 \pm 0.04\,\rm mm/s.
$
Using the same procedure we also measure the rotational diffusivity $D_R$ of the rollers  defined as the exponential decorrelation rate of the velocity orientation in an isotropic phase:
$
D_R = 2.2 \pm 0.1\,\rm  s^{-1}
$.

\section{\label{sec:Numerics}Numerical methods}
\label{Numerics}
 We solve numerically Eqs.~\eqref{eq:toner_tu} using an open source software package, FENICS. It offers a Finite-Element-Method (FEM) platform for solving Partial Differential Equations (PDEs), see Ref.~\cite{Logg2012}. 
We use periodic boundary conditions in a square box of size $L\times L$ for all numerical resolutions, but for the isolated $-1$ defect configuration of Figs.~\ref{Fig:minus_one} and \ref{Fig:shamrock}. In this case  the boundary conditions are defined as follows: $\bm p(x=-L/2,y,t) = {\bm{e}_{x}}$, $\bm p(x=L/2,y,t)= -{\bm{e}_{x}}$, $\bm p(x,y=L/2,t)= {\bm{e}_{y}}$, $\bm p(x,y=-L/2,t) = -{\bm{e}_{y}}$.
If not specified otherwise, we initialize all our numerical simulations with a homogeneous packing fraction  $\rho_0$, and  random velocity fields, and chose hydrodynamic coefficients matching the experimental values measured, and estimated in Refs.~\cite{Geyer2018,Supekar2021}:  $\rho_0=0.1$, $\lambda=0.7$, $\sigma = 5\,\rm mm^{2}\;s^{-2}$, $D=10^{-2}\,\rm mm^{2}\;s^{-1}$, $\alpha = \alpha_{0} (\rho - \rho_{c}) $ where $\alpha_{0}= 10^2\,\rm s^{-1}$ and $\rho_{c} = 3 \times 10^{-3}$, $\beta = 10\,\rm mm^{-2}\;s$. All densities are normalized by $1/(\pi a^2)$ where $a=2.4\,\rm \mu m$ is a colloid radius. $L$ is varied from $1\,\rm mm$ to $10\,\rm mm$.

The time step, between two time increments ${\delta}t$ is chosen to be small compared to the typical relaxation timescale of the fast speed mode $\tau= {\alpha^{-1}(\rho_0)}=10^{-1}\,\rm s$. In practice we take $5 \times 10^{-4}\,s<\delta t \leq 5\times10^{-3}\,s$. The computational mesh consists of $2N\times N$ triangular cells.  The number of triangles per unit length is constant in all the simulations: $N=L/\delta L$, with $\delta L=0.09\,\rm mm$. The density and velocity fields are interpolated using second order polynomials on Lagrange finite element cells, see Ref.~\cite{Logg2012}.

\section{\label{sec:vortex_appendix}Vortex solution for $+1$ defects}
In this Appendix, we provide a simple derivation of the steady-state solution of Toner-Tu equations around an isolated $+1$ defect located at the origin. Assuming that in steady state the Ginzburg-Landau term of Eq.~(\ref{eq:toner_tu}b) is saturated, i.e.  $v^{2}=v_{0}^{2}(1-\rho_{\rm c}/\rho)$,  we seek for a vortex solution of Eq. \eqref{eq:force_balance} of the form $\bm{v}=v\bm{e}_{\phi}$, with $v=v(r)$ by virtue of the azimuthal symmetry of the flow ($\bm{e}_{\phi}$ indicates the azimuthal direction). With this ansatz, the velocity gradient tensor in Eq. \eqref{eq:force_balance} can be expressed as:
\begin{equation}
\label{eq:velocity_gradient}
\nabla\bm{v} = \partial_{r}v\,\bm{e}_{r}\bm{e}_{\phi}-\frac{v}{r}\,\bm{e}_{\phi}\bm{e}_{r}\;,	
\end{equation}
where $\bm{e}_{r}$ is the radial unit vector. Next, replacing Eq. \eqref{eq:velocity_gradient} in Eq. 
\eqref{eq:force_balance}, one finds
\begin{equation}
r\,\partial_{r}\rho=\frac{\lambda v_{0}^{2}}{\sigma}\left(1-\frac{\rho_{\rm c}}{\rho}\right)\;,
\end{equation}
whose solution can be expressed in terms of the Lambert function $W=W(z)$ defined as  the solution of the transcendental equation $W\exp(W) = z$ (see e.g. Ref. \cite{Veberivc2012}), and which satisfies
\begin{equation}\label{eq:lambert}
z\,\partial_{z}W = \frac{W}{1+W}\;.	
\end{equation}
Setting $z=r^{\Lambda}$, $W=\rho/\rho_{\rm c}-1$ and using Eq. \eqref{eq:lambert} finally yields:
\begin{subequations}\label{eq:vortex}
\begin{gather}
\rho= \rho_{\rm c}\left\{1+W\left[\left(\frac{r}{a}\right)^{\Lambda}\right]\right\}\;,\\[5pt]
\bm{v} = v_{0}\sqrt{1-\frac{\rho}{\rho_{\rm c}}}\;\bm{e}_{\phi}\;,
\end{gather}
\end{subequations}
where $\Lambda=\lambda v_{0}^{2}/(\sigma\rho_{\rm c})$. Importantly, the integration constant $a$ represents the core radius of the vortex, namely the distance below which the order parameter is significantly lower than its preferred value: i.e. $v\ll v_{0}$. Unlike in liquid crystals, however, here there is no intrinsic length scale setting this quantity, which is then determined by the geometry of the system as well as the position of other defects \cite{Bricard2015}. 

 Since $W(z) \approx z$, for $0\le z \ll 1$ and $W(z) \approx \log(z)$, for $z \gg 1$, one can readily find asymptotic expansions for the density field $\rho$ in near- and far-field of the vortex, namely:
\begin{equation}
\rho
\approx
\rho_{\rm c}
\left\{
\begin{array}{lll}
1+\left(\frac{r}{a}\right)^{\Lambda} & & r \approx a\\[5pt]
1+\Lambda\log\left(\frac{r}{a}\right) & & r\gg a
\end{array}
\right.\;.
\label{eq:asymptotics}
\end{equation}
It is worth noticing that, opposite to point vortices in inviscid fluids, whose velocity field $\bm{v}=\bm{e}_{\phi}/(2\pi r)$ monotonically decays with the distance $r$ from the center, in flocking matter the speed of a vortex increases with $r$ and eventually saturate at large distances, where $\rho \gg \rho_{\rm c}$ and $v$ reaches the maximal speed of the self-propelled units. 

In practice we use Eq.~\eqref{eq:asymptotics} in Supplemental Material to measure the ratio between the two material parameters $\sigma/\lambda$.

\section{\label{sec:boundary_layer_appendix}Boundary layer solution for $-1$ defects}
The inhomogeneity in the speed of the flow around $-1$ defects cannot be entirely removed, but it can nevertheless be partially relieved by focusing density variations along extended, but narrow, domain walls emanating from the defect center and acting as interfaces between homogeneous portions of the fluid.
Here, we detail the structure of the domain walls and their stabilizing effect. To do so, we construct an approximated piece-wise uniform solution of Eqs.~\eqref{eq:toner_tu}. Orienting the system as illustrated in Fig.~\ref{Fig:shamrock}(a) and labeling with $\rho_{\lessgtr}$ the largest and smallest density value attained by the system, and such that $\rho_{0}=(\rho_{<}+\rho_{>})/2$, the solution takes the form:
\begin{equation}\label{eq:bow_tie}
(\rho,\bm{v}) 
= \left\{
\begin{array}{lll}
(\rho_{>},\,\pm v_{>}\bm{e}_{x}) & & \text{East/West}\\[5pt]
(\rho_{<},\,\mp v_{<}\bm{e}_{y}) & & \text{North/South}\;,
\end{array}
\right.
\end{equation}
where $v_{\lessgtr}^{2}=v_{0}^{2}(1-\rho_{\rm c}/\rho_{\lessgtr})$ and East/West/North/South denotes the angular position of the four wedge-shaped regions, as displayed in Fig.~\ref{Fig:shamrock}(a). Adjacent regions have, in general, different shape, but equal area, being delimited by isosceles triangles with a common edge. Evidently, Eq.~\eqref{eq:bow_tie} is a bulk solution of Eqs.~ \eqref{eq:toner_tu}, featuring a $-1$ defect at the origin. 

The problem we want to solve reduces to finding the constants $\rho_{\lessgtr}$ as well as the configuration of the both fields within the domain walls. To address the first task we seek for a weak solution of Eq. (\ref{eq:toner_tu}a). The latter is obtained by expressing with
\begin{subequations}
\begin{align}
\hspace{3ex}\bm{V}_{\rm in} &\hspace{0.5ex}=\hspace{0.5ex} v_{<}(\cos\theta_{\rm in}\,\bm{N}+\sin\theta_{\rm in}\,\bm{T})\;,\\
\hspace{3ex}\bm{V}_{\rm out} &= v_{>}(\cos\theta_{\rm out}\,\bm{N}+\sin\theta_{\rm out}\,\bm{T})\;,
\end{align}
\end{subequations}
the velocity of the streamlines entering and exiting the domain wall on the basis of the normal and tangent vector, i.e. $\{\bm{N},\bm{T}\}$ [see Fig.~\ref{Fig:shamrock}(a)]. Then, assuming $\rho$ stationary and integrating both sides of Eq.~ (\ref{eq:toner_tu}a) in a rectangular domain spanning a segment of the domain wall, gives
\begin{equation}\label{eq:snell_law}
0 = \int {\rm d}A\,\nabla\cdot(\rho\bm{v}) = \rho_{<}v_{<} \cos \theta_{\rm in}+\rho_{>}v_{>} \cos\theta_{\rm out}\;.   
\end{equation}
Parameterizing $\rho_{>}=2\varphi\rho_{0}$ and $\rho_{<}=2(1-\varphi)\rho_{0}$, with $1/2 \le\varphi\le 1$ and $\rho_{0}=(\rho_{<}+\rho_{>})/2$, approximating $v_{\lessgtr} \approx v_{0}$, under the assumption that $\rho_{\lessgtr} \gg \rho_{\rm c}$, and solving Eq. \eqref{eq:snell_law} with respect to $\varphi$, yields Eq. \eqref{eq:density_fraction}, where we made use of the fact that $|\theta_{\rm in}|-|\theta_{\rm out}|=\pi/2$.

To compute the configuration of the velocity field in the interior of the domain wall, we focus on the semi-infinite domain above the centerline of the domain wall illustrated in Fig.~\ref{Fig:shamrock}(a). Assuming Eq.~\eqref{eq:force_balance} to hold in the interior of the domain wall, Eq.~(\ref{eq:toner_tu}b) reduces to:
\begin{equation}\label{eq:reduced_toner_tu}
\partial_{t}\bm{v} = D \nabla^{2}\bm{v}+(\alpha-\beta v^{2})\bm{v} = 0\;.
\end{equation}
with boundary conditions:
\begin{equation}
\bm{v}(0) = \bm{0}\;,\qquad \lim_{r\rightarrow\infty}\bm{v} = \bm{V}_{\rm out}\;.	
\end{equation}
Now, following the discussion of Sec.~\ref{sec:bow_tie}, we assume the streamlines to be straight, so that, in the internal upper half of the domain wall $\bm{v}$ is parallel to $\bm{V}_{\rm out}$ and its magnitude is a function of the sole arc-length distance $s$ along the streamline. Under these assumptions, using Eq. \eqref{eq:boundary_layer_solution} and expanding Eq. \eqref{eq:reduced_toner_tu} at the cubic order in $v/v_{>}<1$ yields
\begin{equation}\label{eq:boundary_layer}
D\partial_{\parallel}^{2}v+(\alpha_{>}-\beta_{>} v^{2})v = 0\;,
\end{equation}
with the renormalized mean-field coefficients
\begin{subequations}
\begin{gather}
\alpha_{>} = \alpha_{0}\left(\rho_{>}-\rho_{\rm c}+\frac{\lambda}{2\sigma}\,v_{>}^{2}\right)\;,\\
\beta_{>} = \beta_{0}\left[\rho_{>}+\frac{\lambda}{2\sigma}\,\left(v_{0}^{2}+v_{>}^{2}\right)\right]\;.
\end{gather}	
\end{subequations}
Finally, solving Eq. \eqref{eq:boundary_layer} yields
\begin{equation}
v 
= \sqrt{\frac{\alpha_{>}}{\beta_{>}}}\,\tanh\left(\frac{s}{\sqrt{2D/\alpha_{>}}}\right)\;,
\end{equation}
where length scale $\sqrt{2D/\alpha_{>}}$ sets the upper half-width of the domain wall. Using the same algebraic manipulations, one can obtain an analogous approximated solution in the lower half of the domain wall, hence Eq. \eqref{eq:boundary_layer_solution}. Finally, in our experiments and, in general, away from the onset of the flocking transition, $v \approx v_{0}$, from which we obtain Eq. (\ref{eq:boundary_layer_solution}b).

\section{Magnus force for $\pm 1$ defects}
To compute the Magnus force $\bm{F}_{\rm M}$, we assume the velocity field to be fully relaxed and Eqs.~\eqref{eq:toner_tu} can be simplified in the form
\begin{gather}\label{eq:toner_tu_simplified}
\partial_{t}\rho + \nabla\cdot(\rho\bm{v}) = 0\;,\\
\partial_{t}\bm{v} + \lambda\bm{v}\cdot\nabla\bm{v}+\sigma\nabla\rho = \bm{0}\;.
\end{gather}
The force acting along an arbitrary contour $\mathcal{C}$ of the system, thus in particular on the boundary of the defect core, can then be expressed as
\begin{equation}
\bm{F}_{\rm M} = -\oint_{\mathcal{C}} {\rm d}\ell\,\bm{\Pi}\cdot\bm{N}\;,
\end{equation}
where $\bm{\Pi}$ is the momentum current density defined from the equation
\begin{equation}\label{eq:momentum_eom}
\partial_{t}(\rho\bm{v}) + \nabla\cdot\bm{\Pi} = \bm{0}\;.
\end{equation}
Using Eqs. (\ref{eq:toner_tu_simplified}b) one can find, after standard algebraic manipulations
\begin{equation}\label{eq:momentum_flux}
\partial_{t}(\rho\bm{v})
+ \nabla\cdot\left(\frac{1}{2}\,\sigma\rho^{2}\mathbb{1}
+ \lambda\rho\bm{v}\bm{v}\right)
= (1-\lambda)\bm{v}\partial_{t}\rho\;.
\end{equation}
Now, as the motion of the core generally occurs at a much slower rate compared to the relaxation of the fields $\rho$ and $\rho\bm{v}$, one can assume all time derivatives to vanish, expect for $\dot{\bm{R}}={\rm d}\bm{R}/{\rm d}t$, with $\bm{R}$ the position of the core's center. Using Eqs. \eqref{eq:momentum_eom} and \eqref{eq:momentum_flux} then yields
\begin{equation}\label{eq:stress_tensor}
\bm{\Pi} = \frac{1}{2}\,\sigma\rho^{2}\mathbb{1}+\lambda\rho\bm{v}\bm{v}\;,
\end{equation}
as well as 
\begin{equation}\label{eq:stress_trace}
\Pi 
= \sigma\rho^{2}+\lambda\rho v^{2}
= {\rm const}\;,
\end{equation}
where $\Pi=\tr(\bm{\Pi})$ and 
using the fact that $\rho\bm{v}$ is approximatively constant at the time scale of a moving defect. Solving Eq.~\eqref{eq:stress_trace} with respect to $\sigma\rho^{2}$, replacing this in Eq. \eqref{eq:stress_tensor} and expressing the result the in the reference frame of the moving defect, gives
\begin{equation}
\bm{\Pi} 
= \frac{1}{2}\left(\Pi-\lambda\rho|\bm{v}-\dot{\bm{R}}|^{2}\right)\mathbb{1}
+ \lambda\rho(\bm{v}-\dot{\bm{R}})(\bm{v}-\dot{\bm{R}})\;.	
\end{equation} 
Dotting $\bm{\Pi}$ with the normal vector $\bm{N}$ and integrating over the contour encircling the defect core, thus yields two contributions:
\begin{multline}\label{eq:magnus_derivation}
\bm{F}_{\rm M} 
= \frac{1}{2}\lambda\oint_{\mathcal{C}} {\rm d}\ell\,\rho|\bm{v}-\dot{\bm{R}}|^{2}\bm{N}\\	
- \lambda\oint_{\mathcal{C}}{\rm d}\ell\,\rho(\bm{v}-\dot{\bm{R}})(\bm{v}-\dot{\bm{R}})\cdot\bm{N}
\end{multline}
where we assumed that $\oint_{\mathcal{C}} {\rm d}\ell\,\rho\bm{N}=\bm{0}$ because of the symmetric structure of the core. 
Next, expressing $\bm{v}=\bm{V}+\bm{V}_{\rm far}$, with $\bm{V}$ the contribution to the flow velocity originating from the moving defect, and $\bm{V}_{\rm far}$ any far-field contribution possibly caused by other defects, we can compute the integrals at the right-hand side of Eq. \eqref{eq:magnus_derivation}, which yields
\begin{equation}
\bm{F}_{\rm M} 
= \bm{\Gamma}\cdot(\bm{V}_{\rm far}-\dot{\bm{R}})\;,
\end{equation}
where the effective (transverse) drag tensor takes the general form
\begin{equation}
\bm{\Gamma} 
= \lambda\oint_{\mathcal{C}}{\rm d}\ell\,\rho(\bm{V}\bm{N}-\bm{N}\bm{V})\;.
\end{equation}
In deriving this equation we assumed that $\bm{V}_{\rm far}$ and $\dot{\bm{R}}$ are approximately uniform within the core and that, consistently with experimental evidence, $\oint_{\mathcal{C}}{\rm d}\ell\,\rho\bm{V}\cdot\bm{N}=0$ for both $\pm 1$ defects. Finally, expressing $\bm{V}$ in the basis $\{\bm{N},\bm{T}\}$ of the normal and vector one can explicitly calculate the integral and find $\bm{\Gamma}=\Gamma\bm{\epsilon}$, with $\Gamma$ given in Eq.~\eqref{eq:gamma}, where ${\rm d}\bm{\ell}={\rm d}\ell\,\bm{T}$.

\bibliography{Biblio}
\end{document}


\title{
SUPPLEMENTARY INFORMATION:\\~\\
Topology-driven ordering of flocking matter
}
\author{Am\'elie Chardac$^1$}
\author{Ludwig A. Hoffmann$^2$}
\author{Yoann Poupart$^1$}
\author{Luca Giomi$^2$}
\author{Denis Bartolo$^1$}
\affiliation{$^1$Univ.~Lyon, ENS de Lyon, Univ.~Claude Bernard, CNRS, Laboratoire de Physique, F-69342, Lyon, France}
\affiliation{$^2$ Instituut-Lorentz, Universiteit Leiden, P.O. Box 9506, 2300 RA Leiden, The Netherlands}
%

%

\maketitle
\tableofcontents

\newpage

\newpage
\section{Detection of the topological defects}
We detect and measure the topological charge of the  defects of  the polarization field $\bm p (\mathbf{r},t)\equiv(\cos \theta(\mathbf r,t),\sin \theta(\mathbf r,t))$ as follows.
We compute the winding number $w_{ij}$ of $\bm p$ at each point $(i,j)$ on the PIV grid as sketched in Fig.~\ref{SI_Defects}(b). In practice, $w_{ij}$ is the sum of the differences between the adjacent $\theta$ angles along the contour defined by the 8 nearest neighbors of $(i,j)$. 
In the absence of a singularity $w_{ij}$ vanishes. Conversely, at a singular point, $w_{ij}$ takes a finite and quantized value defining the charge of the defect $q=w_{ij}/(2\pi)\in\mathbb Z$. In fact, we only observe $\pm 1$ defects, see Figs.~\ref{SI_Defects}(a) and (c). 
 
\begin{figure} [h!]
	\begin{center}
		\includegraphics[width=0.95\textwidth]{FigSI/SI_Defects.pdf} 
		\caption{{\bf Detection of the topological defects}. 
    \textbf{(a)} Instantaneous polarization field during the coarsening process of a polar flock confined in a $\rm 7\,mm$ wide circular chamber ($\rm t=9.5\,s$). The flow field is marred with $+1$ and $-1$ defects.
    \textbf{(b)} Calculation of the local winding $w_{ij}$ of the polarization field.
    \textbf{(c)} Polarization field with detected topological defects. Gray square indicates the PIV box size ($83.2\,\rm \mu m \times 83.2\,\rm \mu m$).
     }
		\label{SI_Defects}
		\end{center}
\end{figure}

We estimate the accuracy of our detection algorithm by using the conservation of the total topological charge equals $\mathcal{C}=+1$ in our circular chamber where the polarization field is tangent to the boundary. At time $t$, it is given by the difference between the number of $\pm 1$ defects: $\mathcal{C}(t)=\mathcal{N}_{+}(t)-\mathcal{N}_{-}(t)$. The error on the defect count is therefore estimated by $\delta \mathcal N_\pm \sim \frac{1}{\sqrt{2}}(\mathcal C(t)-1)$. This quantity is smaller than 2. The error bars are therefore smaller that the symbol size in all plots of $\mathcal N(t)=\mathcal{N}_{+}(t)+\mathcal{N}_{-}(t)$.

\clearpage
\section{Comparison between Toner-Tu and Ginzburg-Landau coarsenings}
\begin{figure*}[h!]
	\begin{center}
		\includegraphics[width=\textwidth]{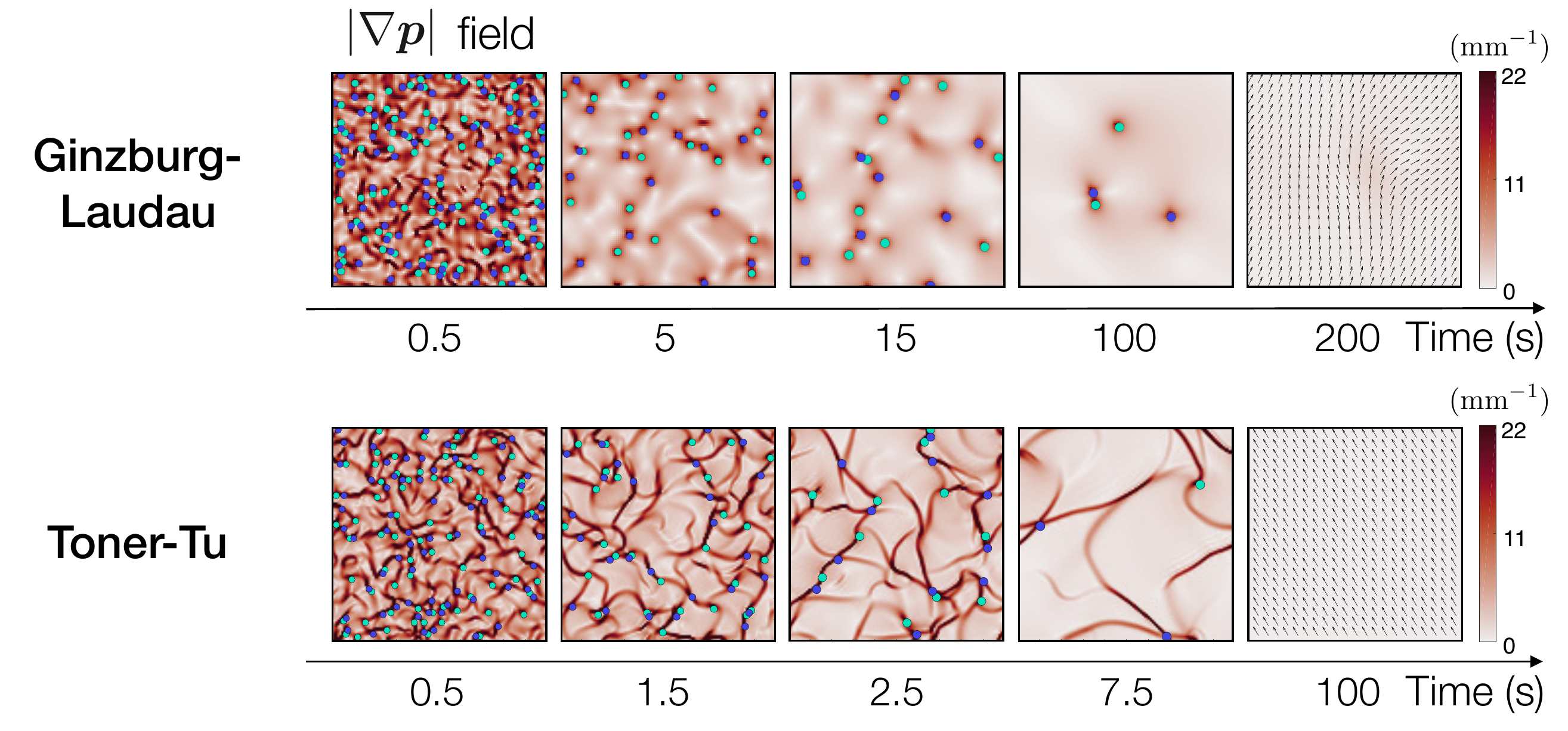} 
		\caption{{\bf Numerical resolution of Ginzburg-Laudau and Toner-Tu equations.} Plotting the polarization strain field, we see that the network of singularity lines seen in flocking liquids (bottom row) is an active property absent of the coarsening dynamics of a passive polar system (top row). Same hydrodynamic parameters as in Fig.~1 in the main text. Both numerical resolutions are computed in a square of size $5\,\rm mm$ with periodic boundary conditions.
		}
		\label{GL}
		\end{center}
\end{figure*}
We recall the Toner-Tu hydrodynamic equation (Eqs.~(1) in the main text):
\begin{subequations}
\label{eq:toner_tu}
\begin{gather}
\partial_{t}\rho+\nabla\cdot(\rho\bm{v}) = 0\;,\\
\partial_{t}\bm{v}+\lambda\bm{v}\cdot\nabla\bm{v} = (\alpha-\beta v^{2})\bm{v} + D\nabla^{2}\bm{v} -
\sigma\nabla\rho
\end{gather}
\end{subequations}
In Fig.~\ref{GL}, we compare the coarsening of a flocking liquid described by Eqs.~(\ref{eq:toner_tu}) with the coarsening of a passive polarization field evolving according to the celebrated Ginzburg-Landau dynamics. The Ginzburg-Landau equation corresponds to the limit  $\lambda=\sigma=0$ in Eqs.~(\ref{eq:toner_tu}). Plotting the polarization strain field, we see that the network of singularity lines seen in flocking liquids (bottom row) is an active property absent of the coarsening dynamics of a passive polar system (top row). The sole singularities of the Ginzburg-Landau equation are isolated pointwise defects whereas they are connected by polarization walls in a Toner-Tu fluid. We used the same hydrodynamic parameters as in Fig.~1 of the main text.

\newpage
\section{Impact of the  hydrodynamic coefficients on the phase ordering kinetics}
\begin{figure*}[h!]
	\begin{center}
		\includegraphics[width=\textwidth]{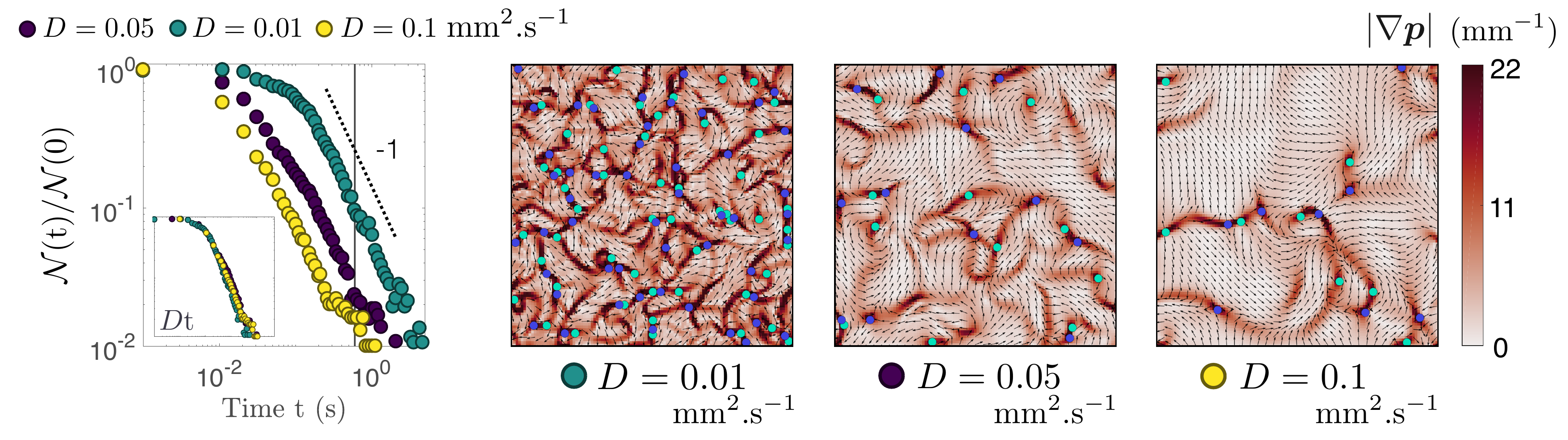} 
		\caption{{\bf Impact of $D$ on the coarsening dynamics. } Left: Time variations of the total number of topological defects for three values of $D$. Right: Snapshots of the polarization strain at time $t= \rm 0.6\,s$ for $D=0.01,\,0.05, \,0.1\,\rm mm^2\cdot s^{-1}$. $\lambda=0.7$, $\alpha_0=100\,\rm s^{-1}$, $\beta=10\,\rm mm^{-2}\cdot s$, $\sigma=5\,\rm mm^{2}\cdot s^{-2}$. 
		}
		\label{DComp}
		\end{center}
\end{figure*}

\begin{figure*}[h!]
	\begin{center}
		\includegraphics[width=\textwidth]{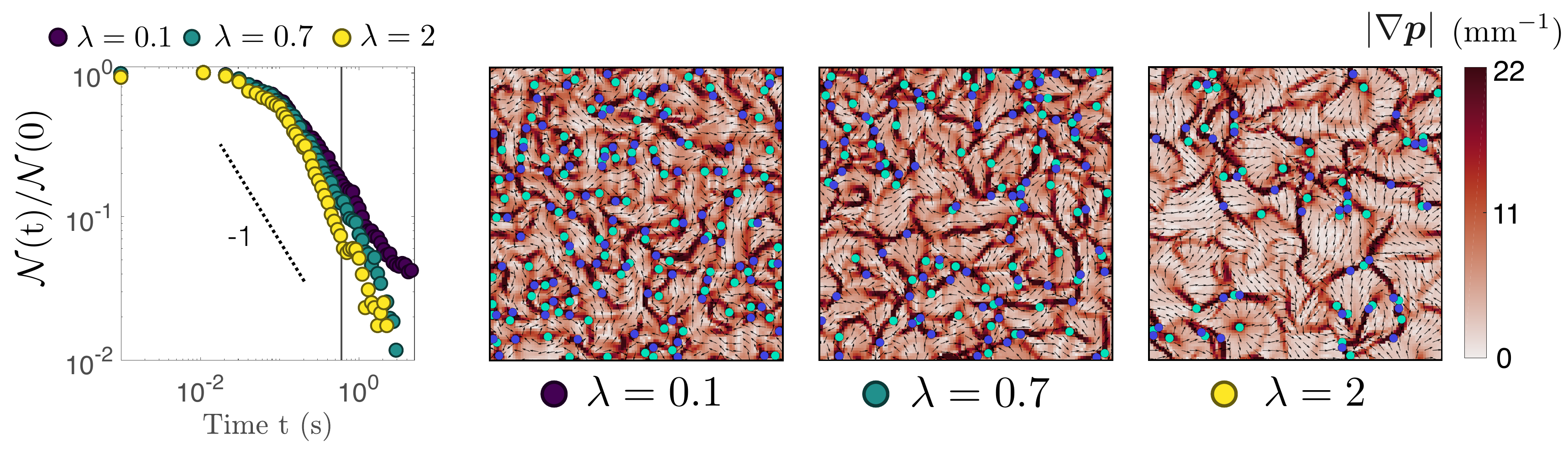} 
		\caption{{\bf Impact of $\lambda$ on the coarsening dynamics. } Left: Time variations of the total number of topological defects for three values of $\lambda$. Right: Snapshots of the polarization strain at time $t= \rm 0.6\,s$ for $\lambda=0.1,\,0.7, \,2$.  $D=0.01\,\rm mm^2\cdot s^{-1}$, $\alpha_0=100\,\rm s^{-1}$, $\beta=10\,\rm mm^{-2}\cdot s$, $\sigma=5\,\rm mm^{2}\cdot s^{-2}$.
		}
		\label{LambdaComp}
		\end{center}
\end{figure*}

\begin{figure*}[h!]
	\begin{center}
		\includegraphics[width=\textwidth]{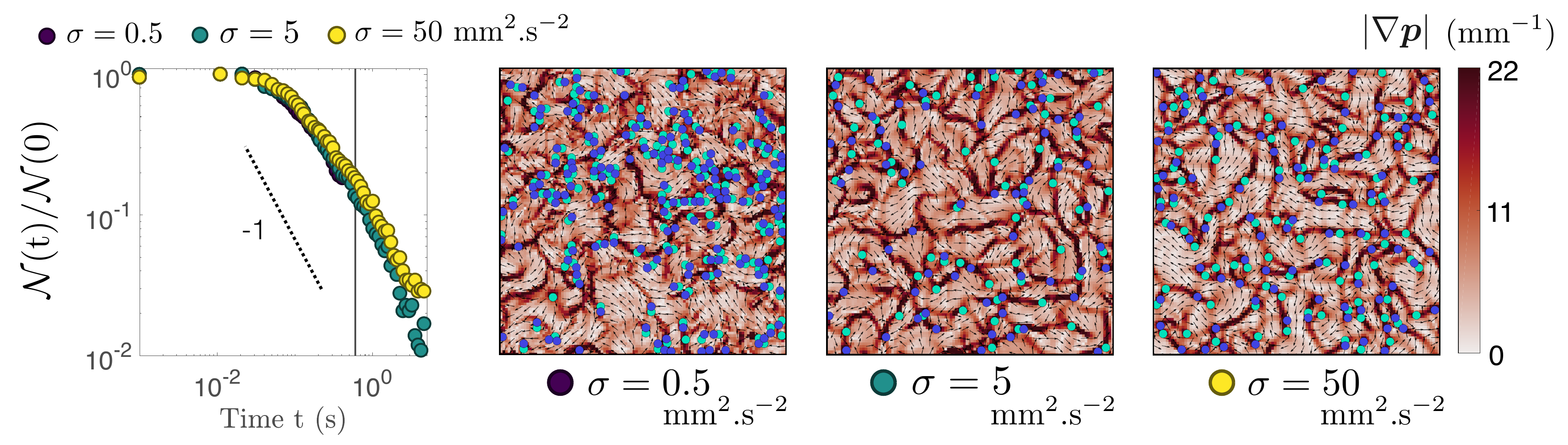} 
		\caption{{\bf Impact of $\sigma$ on the coarsening dynamics. } Left: Time variations of the total number of topological defects for three values of $\sigma$. Right: Snapshots of the polarization strain at time $t= \rm 0.6\,s$ for $\sigma=0.5,\,5, \,50\, \rm mm^2\cdot s^{-2}$. $\lambda=0.7$, $D=0.01\,\rm mm^2\cdot s^{-1}$, $\alpha_0=100\,\rm s^{-1}$, $\beta=10\,\rm mm^{-2}\cdot s$.
		}
		\label{SigmaComp}
		\end{center}
\end{figure*}

\begin{figure*}[h!]
	\begin{center}
		\includegraphics[width=\textwidth]{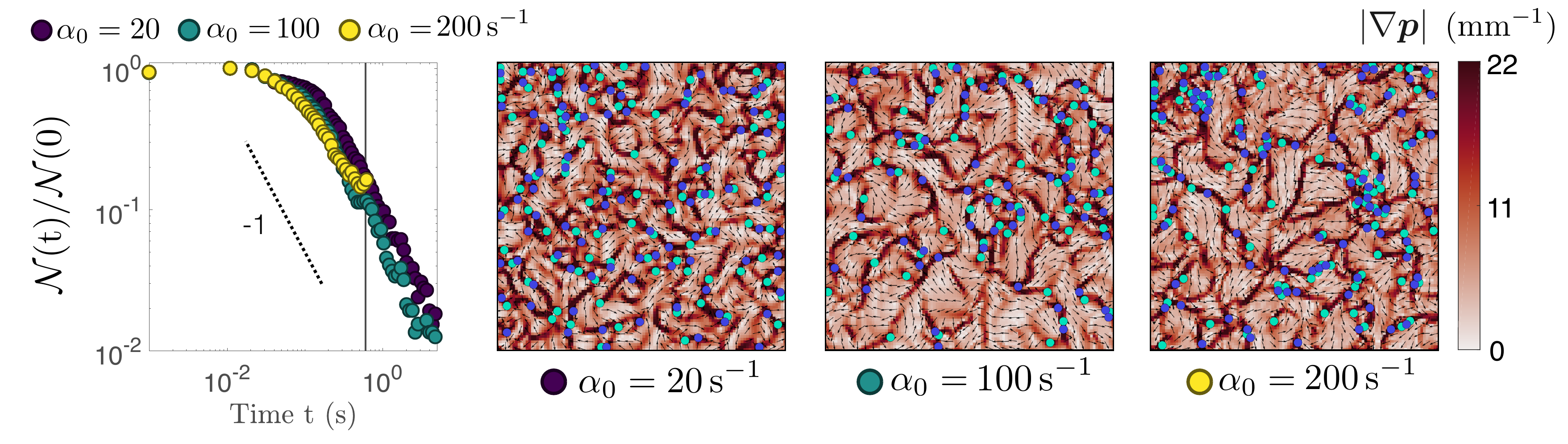} 
		\caption{{\bf Impact of $\alpha_0$ on the coarsening dynamics. } Left: Time variations of the total number of topological defects for three values of $\alpha_0$.   Right: Snapshots of the polarization strain at time $t= \rm 0.6\,s$ for $\alpha_0=20,\,100, \,200\,\rm s^{-1}$. $\lambda=0.7$,  $D=0.01\,\rm mm^2\cdot s^{-1}$,  $\beta=10\,\rm mm^{-2}\cdot s$, $\sigma=5\,\rm mm^{2}\cdot s^{-2}$.
		}
		\label{a2Comp}
		\end{center}
\end{figure*}

\begin{figure*}[h!]
	\begin{center}
	\includegraphics[width=\textwidth]{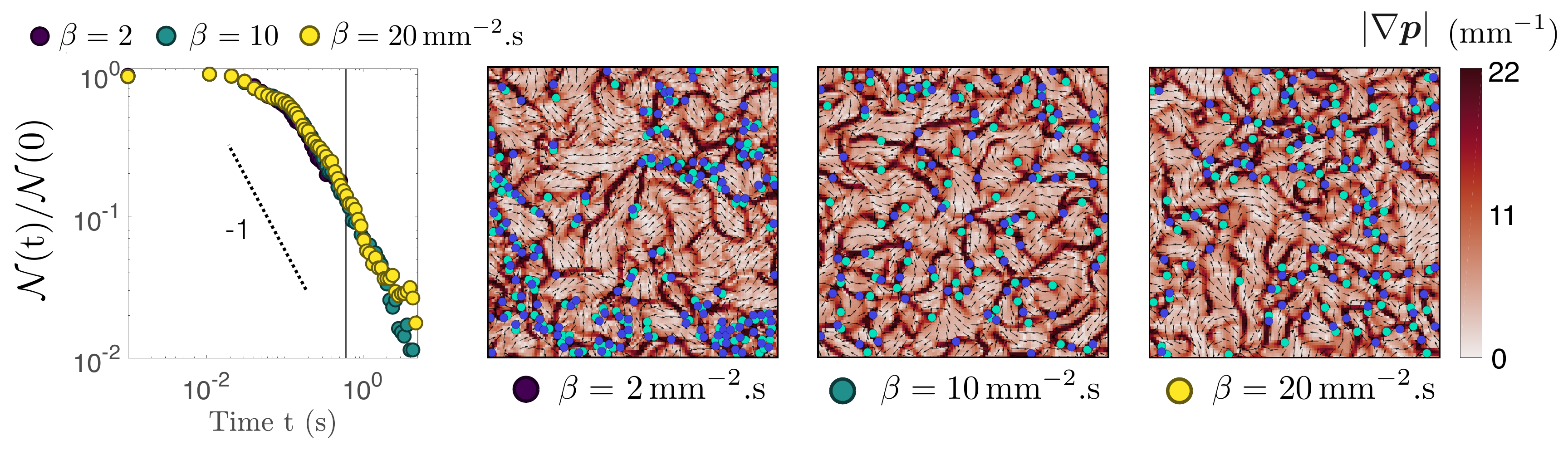} 
		\caption{{\bf Impact of $\beta$ on the coarsening dynamics. } Left: Time variations of the total number of topological defects for three values of $\beta$. Right: Snapshots of the polarization strain at time $t= \rm 0.6\,s$ for $\beta=2,\,10, \,20\,\rm mm^{-2}\cdot s$. $\alpha_0=100 \,\rm s^{-1}$, $\lambda=0.7$, $D=0.01\,\rm mm^2\cdot s^{-1}$, $\sigma=5\,\rm mm^{2}\cdot s^{-2}$.
		}
		\label{a4Comp}
		\end{center}
\end{figure*}

In this section, we take advantage of the numerical resolution of Toner-Tu hydrodynamics to single out the 
physical mechanisms ruling the coarsening dynamics of the polarization field. To do so we solve Eqs.~(\ref{eq:toner_tu}) varying the hydrodynamic coefficients one at a time over an order of magnitude. In agreement with our theory, Figs.~\ref{DComp},~\ref{LambdaComp},~\ref{SigmaComp},~\ref{a2Comp} and~\ref{a4Comp} reveal that the phase ordering kinetics of the flocking fluid is primarily determined by the orientational elasticity of the flock $D$. Varying all the other hydrodynamic parameters hardly alters the emergent polarization wall network and the  defect annihilation dynamics. In the inset of the first panel of Fig.~\ref{DComp}, we clearly see that all the $\mathcal N(t)$ curves collapse on a single mastercurve when plotting them against $Dt$.

\newpage
\section{Modes of deformation}
Our  theory relies on the assumption of a divergence-free active flow. We show below that this simplification is very well supported by our measurements. To do so we decompose the polarization strain tensor $\partial_ip_j$ on the orthogonal basis defined by: 
\begin{gather}
    \rm M_{1} =
\begin{pmatrix} 1 & 0 \\ 0 & 1 \end{pmatrix},\, M_{2} =\begin{pmatrix} 0 & 1 \\ -1 & 0 \end{pmatrix},\, M_{3} =\begin{pmatrix} 1 & 0 \\ 0 & -1 \end{pmatrix},\, and \,M_{4} =\begin{pmatrix} 0 & 1 \\ 1 & 0 \end{pmatrix}.
\end{gather}
We can then write $\mathbf{\nabla} \bm p =d M_{1} + \omega M_{2} + s_{0}M_{3}+s_{45}M_{4} $ where $d$ is the flow divergence, $\omega$ the curl, $s_{0}$ the $x$-axis shear and $s_{45}$ the  pure shear oriented at a $45^\circ$ angle with the $x-$axis.  In Fig.~\ref{SI_Modes} we plot the local magnitude of all the four deformation modes, together with the magnitude of the total strain $|\bm\nabla\bm p|=(d^2+\omega^2+s_0^2+s_{45}^2)^{1/2}$ measured in our experiments. The flow divergence is vanishingly small away from the domain walls where the strain is dominated by the two independent shear modes.

 \begin{figure*}[h!]
	\begin{center}
		\includegraphics[width=\textwidth]{FigSI/SI_ModesDeformation.pdf}
		\caption{{\bf The active flows are nearly divergenceless.} In order to quantify the relative magnitude of the flow divergence compared to the other deformation modes we compare the magnitude of the total strain, of the divergence $d$, of the curl $\omega$ and of the two independent shear modes $s_0$ and $s_{45}$. The flow is divergence-less away from the domain walls where the deformations are primarily decomposed on the two shear modes.  Same experiment as in Fig.~1(c) in the main text at time $\rm t=4\, s$. Scale bar: $1\,\rm mm$.}
		\label{SI_Modes}
		\end{center}
\end{figure*} 

\newpage
\section{Measurements of the structural scales of the domain-wall networks}
\subsubsection{Correlation length of the polarization field}
\begin{figure} [h!]
	\begin{center}
		\includegraphics[width=0.8\textwidth]{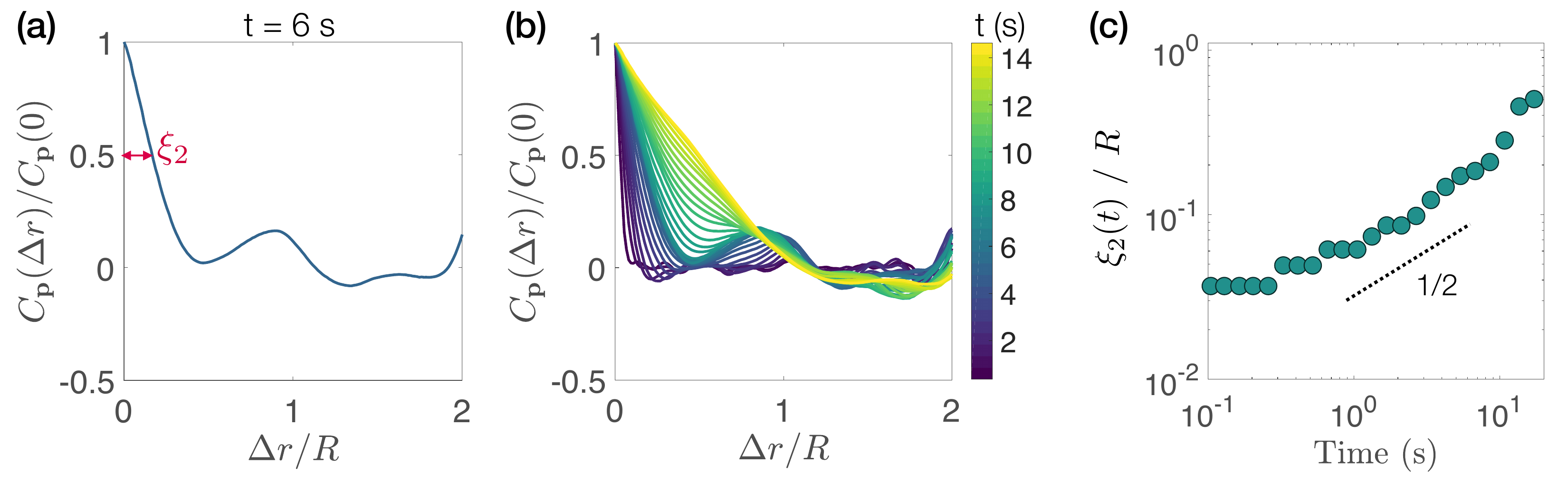}
		\caption{{\bf Correlations of the polarization field.} {\bf (a)} Decay of the correlation of the polarization field measured in the experiment corresponding to Fig.~1 in the main text, measured at time $t=6\,\rm s$ after the onset of the colloid propulsion.  {\bf (b)} Evolution of the correlation function with time. {\bf (c)} The half-height width of the correlations plotted in {\bf (b)} define the correlation length $\xi_2(t)$.}
		\label{SI_Correlation}
		\end{center}
\end{figure}

We measure the orientational correlation length $\xi_2(t)$ from the spatial decay of the (connex) two point function $C_p(R)=\langle \bm p(\mathbf r,t)\cdot \bm p(\mathbf r+\mathbf r,t)\rangle_{r}$, see Figs.\ref{SI_Correlation}(a), \ref{SI_Correlation}(b), \ref{SI_Correlation}(c) .  $\xi_2(t)$ is defined as the half-height width of $C_p(R)$ measured at time $t$. 

\subsubsection{Typical size of the incompatible polarization domains}
\begin{figure*} [h!]
	\begin{center}
		\includegraphics[width=0.7\textwidth]{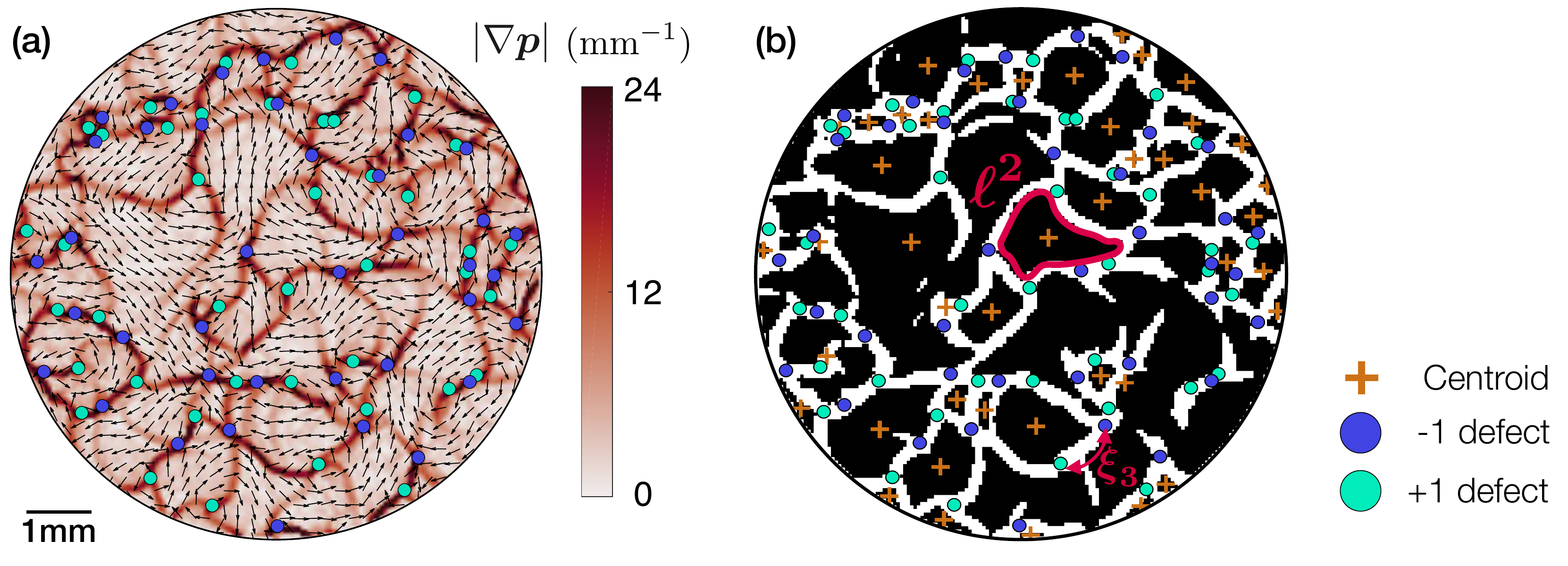} 	\caption{{\bf Geometry of the polarization domains.} {\bf (a)} Map of the instantaneous strain polarization and location of the topological defects (experiments). {\bf (b)} Applying a threshold we isolate the polarization domains separated by domain walls of nearly constant thickness.  From the thresholded maps of strain field, we detect the domains and measure their area. We also measure the line density of defect along the domains (white region) to define $\xi_3$.
     }
		\label{Threshold}
		\end{center}
\end{figure*}

We detail the method used to measure the typical size $\ell(t)$ of the regions of incompatible polarization at time $t$ and the typical curvilinear distance $\xi_{3}$ between the topological defects along the polarization wall network. The method is illustrated on experimental data in Fig.~\ref{Threshold}. The polarization domains are delimited by the plaquettes of the domain-wall network whose the edges are defined by the   regions of highest deformation, Fig.~\ref{Threshold}(a). Applying a threshold to the maps of the polarization strain (Fig.~\ref{Threshold}(b)) and noting that all the domain walls have nearly the same width, we can measure the overall length $L(t)$ of the network edges.  $\xi_3$ is then given by $\xi_3(t)=L(t)/\mathcal N(t)$, where $\mathcal N(t)$ is the number of topological defects at time $t$. 

From the image showed in Fig.~\ref{Threshold}(b), and using the MATLAB function {\em regionprops}, we measure the center of mass of each domain and its area. $\ell^2(t)$ is then defined as the instantaneous average area of the polarization domains.

\subsubsection{Evolution of the domain wall geometry in the experiments and simulations}
\begin{figure*}[h!]
	\begin{center}
		\includegraphics[width=0.95\textwidth]{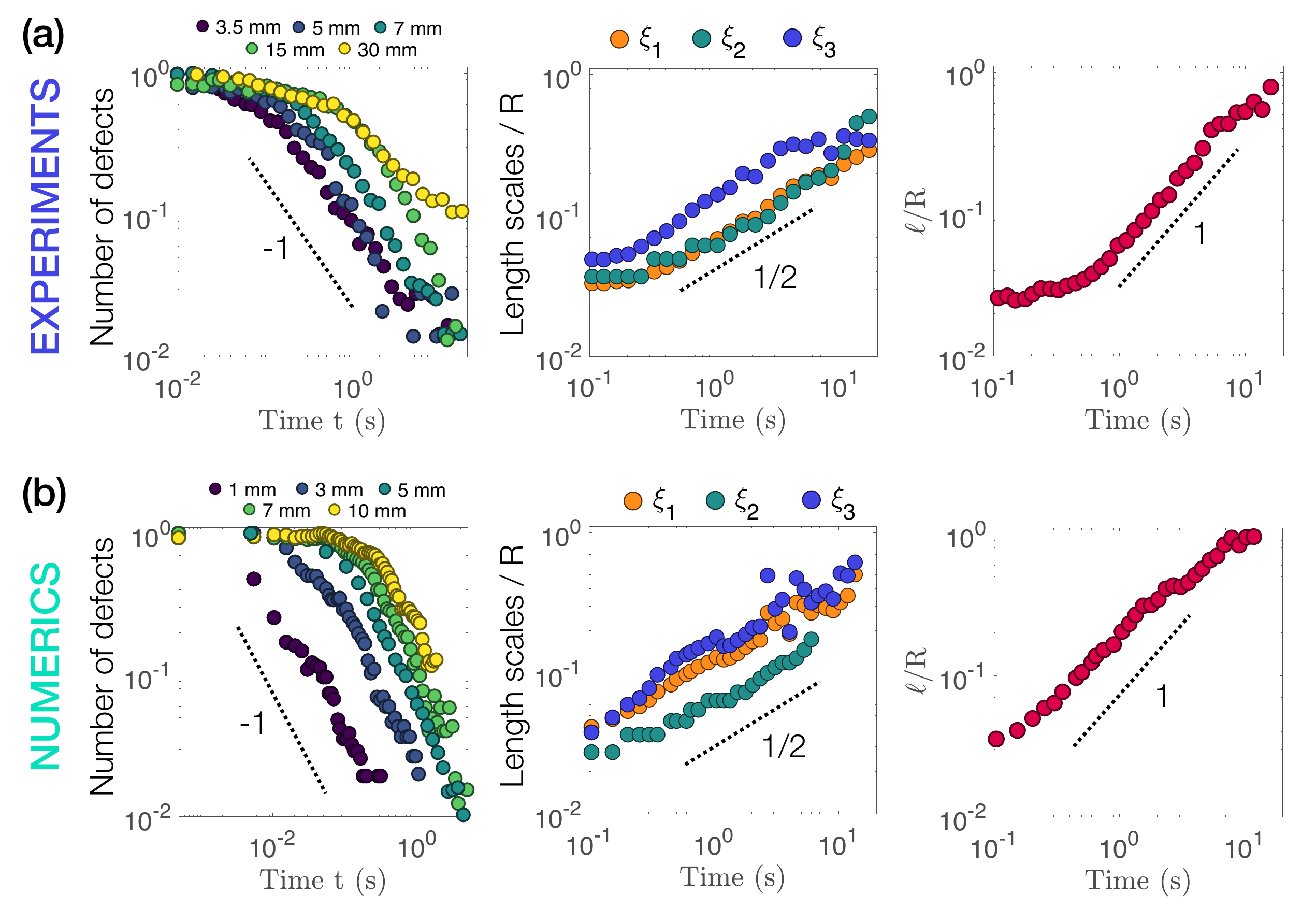} 	\caption{{\bf Structural length scales.} {\bf (a)} Time evolution of the four structural scales $\xi_1$, $\xi_2$, $\xi_3$ and $\ell$ measured in our experiments. Same geometry as in Fig. 1 in the main text.  {\bf (b)} Same quantities measured in our simulations of a model Toner-Tu Fluid. Same simulation parameters as in Fig. 1.
     }
		\label{SI_Correlation_comparaison}
		\end{center}
\end{figure*}

As illustrated in Figs.~\ref{SI_Correlation_comparaison}(a) and \ref{SI_Correlation_comparaison}(b). All structural length scales defined in the main text evolve according to the same scaling law in our experiments and simulations.

\newpage
\section{Estimation of hydrodynamic coefficients from experiments}

To measure the ratio $ \sigma / \lambda$ in our experiments done in the chambers having a shamrock shape, we measure the density and the local kinematic pressure $v^2/r$ in an outer ring illustrated in Fig.~\ref{trefle}(a).  We then plot the ratio $v^2/\partial_r \rho$ as a function of the radial distance, in 
Fig.~\ref{trefle}(c). The slope of the linear fit defines the ratio  $ \sigma / \lambda$ as explained in Appendix C.
\begin{figure*}[h!]
	\begin{center}
		\includegraphics[width=\textwidth]{FigSI/SI_Coeff.pdf} 
		\caption{{\bf Estimation of the ratio $ \sigma / \lambda$.} {\bf (a)} Density field. Scalebar: $1\,\rm mm$. {\bf (b)} Flow speed. Scalebar: $1\,\rm mm$. {\bf (c)} The ratio $v^2/\partial_r \rho$ increases linearly with $r$. A linear fit provides a direct measurement of $ \sigma / \lambda$.
		}
		\label{trefle}
		\end{center}
\end{figure*}

\newpage
\section{Description of the Supplementary Movies}
{\bf Supplementary Movie 1:} Experimental images. Flocking motion emerges in an ensemble of about 4 millions colloidal rollers. The circular chamber has a diameter of $3\,\rm cm$. When the DC electric field is turned on, the rollers start propelling along random directions. After a short transient, they self-organize into a polar fluid. The resulting active flow self-organize in the  form of a steady macroscopic vortex. Colloid area fraction $\rho_{0} = 0.1$. Colloid diameter: $4.8\,\rm \mu m$. Field amplitude: $E_{0} = 8 \,\rm V/\mu m$. Video recorded at 30 fps, played at 300 fps. \\

\textbf{Supplementary Movie 2:} Experimental data. Coarsening dynamics of the polarization field in a $7\,\rm mm$ wide circular chamber. The movie shows both the instantaneous polarization and strain fields. The dots indicate the position of the $\pm 1$ topological defects. Colloid fraction $\rho_{0} = 0.1$. PIV box size: $83.2\,\rm \mu m \times 83.2\,\rm \mu m$. Field amplitude: $E_{0} = 5.2\,\rm V/\mu m$. Video recorded at 200 fps, played at 300 fps. \\

\textbf{Supplementary Movie 3:} Experimental data. Coarsening dynamics of the density field in a $7\,\rm mm$  circular chamber. The movie shows the instantaneous density field. Colloid fraction $\rho_{0} = 0.1$. PIV box size: $83.2\,\rm \mu m \times 83.2\,\rm \mu m$. Field amplitude: $E_{0} = 5.2\,\rm V/\mu m$. Video recorded at 200 fps, played at 300 fps. \\

\textbf{Supplementary Movie 4:} Numerical resolution of Toner-Tu equations. Coarsening dynamics
in a $5\,\rm mm$ wide periodic computational domain. The movie shows the instantaneous polarization field and the corresponding strain field. The dots indicate the position of the $\pm 1$ topological defects. Numerical resolution initialized with a homogeneous colloidal fraction ($\rho_{0} = 0.1$) and random velocity. \\

\textbf{Supplementary Movie 5:} Numerical resolution of Toner-Tu equations. Coarsening dynamics in a $5\,\rm mm$ wide periodic domain. The movie shows the instantaneous density field. Numerical resolution initialized with a homogeneous colloidal fraction ($\rho_{0} = 0.1$) and random velocity. \\

\textbf{Supplementary Movie 6:} Numerical resolution of Ginzburg-Landau equation. Coarsening dynamics of the polarization field in a $5\,\rm mm$ wide periodic domain. The movie shows the instantaneous polarization  and  strain fields. The dots indicate the position of the $\pm 1$ topological defects. Numerical resolution initialized with a homogeneous colloidal fraction ($\rho_{0} = 0.1$) and random velocity. \\
